# Recent progress in exploring magnetocaloric materials

*B. G. Shen,\* J. R. Sun, F. X. Hu, H. W. Zhang,* and *Z. H. Cheng*

State Key Laboratory for Magnetism, Institute of Physics, Chinese Academy of Sciences,

Beijing 100190 (China)

**Abstract**

Magnetic refrigeration based on the magnetocaloric effect (MCE) of materials is a potential technique that has prominet advantages over the currently used gas compression-expansion technique in the sense of its high efficiency and environment friendship. In this article, our recent progress in exploring effective MCE materials is reviewed with the emphasis on the MCE in the $LaFe_{13-x}Si_x$–based alloys with a first order magnetic transition discovered by us. These alloys show large entropy changes in a wide temperature range near room temperature. Effects of magnetic rare-earth doping, interstitial atom, and high pressure on the MCE have been systematically studied. Special issues such as appropriate approaches to determining the MCE associated with the first-order magnetic transition, the depression of magnetic and thermal hystereses, and the key factors determining the magnetic exchange in alloys of this kind are discussed. The applicability of the giant MCE materials to the magnetic refrigeration near ambient temperature is evaluated. A brief review of other materials with significant MCE is also presented in the article.

[*] E-mail: shenbg@g203.iphy.ac.cn

# 1. Introduction

Magnetocaloric effect (MCE) provides a unique way of realizing the refrigeration from ultra-low temperature to room temperature. With the increase of applied field, magnetic entropies decrease and heat is radiated from the magnetic system into the environment through an isothermal process, while with the decrease of applied field, magnetic entropies increase and heat is absorbed from the lattice system into the magnetic system through an adiabatic process. Both the large isothermal entropy change and the adiabatic temperature change characterize the prominent MCE.

The MCE was first discovered by Warburg in 1881.[1] Debye in 1926[2] and Giauque in 1927[3] independently pointed out that ultra-low temperatures could be reached through the reversible temperature change of paramagnetic (PM) salts with the alternation of magnetic field and first foresaw the technological potential of this effect. The first experiment on magnetic refrigeration was performed by Giauque and MacDougall in 1933.[4] With the use of this technology, the temperatures below 1 K were successfully achieved. Nowadays, magnetic refrigeration has become one of the basic technologies getting ultra-low temperatures.

Along with the success in the ultra-low temperature range, this technique in lifted temperature ranges such as ~1.5 K– ~20 K for the range of liquid helium, and ~20 K– ~80 K for the range of liquid hydrogen and liquid nitrogen, is also applied. The mature refrigerants for the rang from ~1.5 K to ~20 K are the garnets of the $R_3M_5O_{12}$ (R=Nd, Gd, and Dy, M=Ga and Al), $Gd_2(SO_4)_3 \cdot 8H_2O$,[3,4] $Dy_3Al_5O_{12}$ (DAG).[5] The most typical material is $Gd_3Ga_5O_{12}$ (GGG),[6] which has been successfully applied to the precooling for the preparation of liquid helium. As for the temperature range ~10 K– ~80 K, the modest refrigerants are Pr, Nd, Er, Tm, $RAl_2$ (R=Er, Ho, Dy and $Dy_{0.5}Ho_{0.5}$, R=$Dy_xEr_{1-x}$ 0<x<1, GdPd ) and $RNi_2$ (R=Gd, Dy and Ho).[7]

Magnetic refrigeration near room temperature is of special interest because of its great social effect and economical benefit. Compared with the refrigeration technology based on the gas-compression/expansion, which has been widely used today, magnetic refrigeration is environment friendly and energy saving. Due to the significant increase of heat capacity near the ambient temperature, the heat transferred by each magnetizing-demagnetizing cycling of the refrigerator should be considerably large to guarantee refrigeration efficiency. As a result, most of the materials working at low temperatures cannot be directly utilized, and new materials with great entropy change around the ambient temperature must be explored. Brown in 1968 observed a large MCE of Gd ($T_C$=293 K).[8] However, the entropy changes of magnetocaloric materials reported subsequently are much smaller than that of Gd. In 1997, it was found by Gschneidner and his colleagues[9] that the entropy change of $Gd_5Si_2Ge_2$ with a fist order phase transition is ~18 J/kgK around $T_C$=278 K for a field change of 0–5 T, which is significantly larger than that of Gd (~10 J/kgK) under a similar condition. Du and his colleagues[10] found that $La_{1-x}Ca_xMnO_3$ (x = 0.2) has a large entropy change of 5.5 J/kgK at 230 K for a field change of 0–1.5 T. Subsequently, some new magnetocaloric materials with a first-order magnetic transition such as $LaFe_{13-x}Si_x$,[11,12] $MnAs_{1-x}Sb_x$,[13] and $MnFeP_{0.45}As_{0.55}$[14] were found to have entropy changes from 18 J/kgK to 30 J/kgK. These achievements aroused a new wave of MCE research.

A common feature of the giant MCE is that it usually occurs being accompanied by a first-order magnetic transition. However, several basic problems involved in the MCE, arising from the first-order phase transition, must be solved, such as, the determination of MCE, the contributions other than magnetic entropy to MCE, and the effects of magnetic/thermal hysteresis as far as practical application being concerned. Detailed reviews of the magnetic refrigeration technology and corresponding materials have been given by Tishin,[15] Gschneidner[16] and Bruck.[17] In the present paper, we give a brief review of the development in exploring magnetocaloric materials, mainly on our recent progress in typical magnetic materials with giant MCE.

## 2. Thermodynamics of magnetocaloric effect

According to the standard thermodynamics,[15] the total differential of Gibbs free energy $G$ of a system under volume $V$, magnetic field $H$, pressure $P$, and at temperature $T$ is given by $dG=VdP-SdT-MdH$. Based on this equation, the well-known Maxwell relation can be obtained as

$$\Delta S(T,H,P) = S(T,H,P) - S(T,H=0,P) = \int_0^H \left(\frac{\partial M}{\partial T}\right)_{H,P} dH \tag{1}$$

if the pressure and temperature are fixed, where $M$ is the magnetization. Equation (1) gives an approach to calculating entropy change from magnetic data. In practice, an alternative formula is usually used for numerical calculation,

$$\Delta S = \sum_i \frac{M_{i+1} - M_i}{T_{i+1} - T_i} \Delta H_i . \tag{2}$$

The following formula is also used:

$$\Delta S(T,H,P) = \int_0^T \frac{(C_{H,P} - C_{0,P})}{T} dT , \tag{3}$$

where $C_{H,P}$ and $C_{0,P}$ are the heat capacities in the fields of $H$ and 0, respectively, under constant pressure $P$. According to the standard thermodynamics, the differential entropy has the following form:

$$dS = \left(\frac{\partial S}{\partial T}\right)_{H,P} dT + \left(\frac{\partial S}{\partial H}\right)_{T,P} dH + \left(\frac{\partial S}{\partial P}\right)_{T,H} dP . \tag{4}$$

$dS=0$ under the adiabatic condition, then the adiabatic temperature change will be

$$dT = -\frac{T}{C_{H,P}} \left(\frac{\partial M}{\partial T}\right)_{H,P} dH , \tag{5}$$

where $C_{H,P}=T(\partial S/\partial T)_{H,P}$ denotes heat capacity. Integrating Eq. (5) over $H$=0-$H$ yields the adiabatic temperature change $\Delta T_{ad}$. This is an indirect technique for the determination of $\Delta T_{ad}$.

## 3. Magnetic and magnetocaloric properties of LaFe$_{13-x}$Si$_x$-based compounds

LaFe$_{13}$ does not exist because of its positive formation enthalpy. To obtain a LaFe$_{13}$-based alloy, a third element has to be introduced. The first stable LaFe$_{13-x}$M$_x$ compound was obtained by Kripyakevich *et al.*[18] in 1968 after partially replacing Fe with Si or Al. The compound crystallizes in the cubic NaZn$_{13}$-type structure with a space group of $Fm\bar{3}c$. The incorporation of Si/Al causes no change in crystal structure but lattice shrinkages because of the small atomic size of Si/Al compared with that of Fe. We measured the lattice constant for LaFe$_{13-x}$Si$_x$ as shown in Figure 1a. The lattice constant was found to decrease with Si content increasing from ~1.1475 nm for x=1.2 to ~1.1450 nm for x=2.5. Although the change is not very large, it has dramatic effects on magnetic property.

The study showed that LaFe$_{13-x}$Si$_x$ is PM near the ambient temperature, and undergoes an FM transition upon cooling at a temperature between 200 and 250 K, depending on Si content. The Curie temperature $T_C$ obtained from our measurements increases linearly from ~175 K for x=1.17 to ~254 K for x=2.5 (Fig. 1a) in conformity with the result reported by Palstra *et al.*[19] We and Palstra *et al.*[19] also observed a monotonic decrease of saturation magnetization, respectively, at a rate of −0.286 μ$_B$ for each Si, with the increase of Si content (Fig. 1b). This is a rate much more rapid than that expected for a simple dilution effect, and therefore indicating the change of electronic structure due to the incorporation of Si. A first-principles calculation conducted by Wang *et al.*[20] indicated the occurrence of hybridization between the Fe-3d and Si-2p electrons and the change of the density of state below Fermi surface after the introduction of Si, which could be the cause for the change of Fe magnetic moment.

### 3.1 Magnetocaloric effect in LaFe$_{13-x}$Si$_x$

The study by Palstra et al.[21] indicated that LaFe$_{13-x}$Si$_x$ compounds were stable in the cubic NaZn$_{13}$ structure in a concentration region 1.5 ≤ x ≤ 2.5. We found that Si content could be further lowered down to x=1.2 through improving preparation arts, while the cubic NaZn$_{13}$ structure remained unchanged. Accompanied with the FM to PM transition, the LaFe$_{13-x}$Si$_x$ compounds exhibit an obvious lattice contraction, especially for the samples with low Si content. Sizable thermal or magnetic hysteresis was observed for the temperature/magnetic field ascending-descending cycling. These results reveal the first-order nature of the phase transition. An evolution of the phase transition from first-order to second-order takes place as x increases, and a typical second-order transition appears when x > 1.8. For LaFe$_{13-x}$Si$_x$ compounds with Si content x ≤ 1.6, an external field can induce metamagnetic transition from PM to FM state at temperatures near but above $T_C$, which is the so-called itinerant-electron metamagnetic (IEM) transition.[22,23] The FM state becomes more stable than the PM state under an applied field due to the field-induced change in the band structure of 3d electrons. The IEM transition is usually indicated by the appearance of "S"-shaped $M^2$-H/M isotherms (Arrott plot). For the LaFe$_{13}$-based compounds, the lower the Si content is, the stronger the first-order nature of the magnetic transition will be.[24]

Although the magnetic property of LaFe$_{13-x}$Si$_x$ had been intensively studied before, the magnetocaloric property did not come into the vision of people until the work by Hu et al.[11] in 2000, they observed an entropy change as high as ~20 J/kgK, for a field change of 0–5 T, in LaFe$_{13-x}$Si$_x$ with lower Si concentration. In 2001 further study by Hu et al.[12] found that the large entropy change |ΔS| in LaFe$_{13-x}$Si$_x$ is associated with negative lattice expansion and metamagnetic transition behaviour. Figure 2a shows the magnetization isotherms of LaFe$_{11.4}$Si$_{1.6}$ as field increases and decreases at different temperatures around its Curie temperature $T_C$=208 K. It is evident that each isotherm shows a reversible behaviour. Almost no temperature and magnetic hysteresis accompanied with the magnetic transition is observed, although the Si content of x=1.6 is located near the critical boundary from first-order to second-order transition. The completely reversible magnetization indicates that |ΔS| should be fully reversible on temperature and magnetic field. Figure 2b exhibits Arrott plots of LaFe$_{11.4}$Si$_{1.6}$. The existence of the inflection point confirms the occurrence of a metamagnetic transition from the paramagnetic to ferromagnetic ordering above the $T_C$. The ΔS of LaF$_{11.4}$Si$_{1.6}$ as a function of temperature is shown in Figure 2c and the maximal values of |ΔS| under fields of 1, 2, and 5 T are 10.5, 14.3, and 19.4 J/kg K, respectively. Another interesting feature is that the ΔS peak broadens asymmetrically to higher temperature with field increasing. Detailed analysis indicates that the field-induced metamagnetic transition above $T_C$ contributes to the asymmetrical broadening of ΔS.

The x-ray diffraction measurements at different temperatures reveal that the LaFe$_{11.4}$Si$_{1.6}$ remains cubic NaZn$_{13}$-type structure but the cell parameter changes dramatically with $T_C$. The negative expansion of lattice parameter reaches 0.4% with appearance of FM ordering for LaFe$_{11.4}$Si$_{1.6}$, while only a small change of lattice parameter is observed for LaFe$_{10.4}$Si$_{2.6}$ (see Fig. 3a). The occurrence of the large |ΔS| in LaFe$_{11.4}$Si$_{1.6}$ is attributed to the rapid change of magnetization at $T_C$, which is caused by a dramatic negative lattice expansion. For comparison, Figure 2c also presents an entropy change of LaFe$_{10.4}$Si$_{2.6}$, and its value is much smaller than that of LaFe$_{11.4}$Si$_{1.6}$. The saturation magnetizations of LaFe$_{11.4}$Si$_{1.6}$ and LaFe$_{10.4}$Si$_{2.6}$ were determined to be 2.1 and 1.9 $\mu_B$/Fe from the *M*–*H* curves at 5 K. The influence of the small difference in saturation magnetization between the two samples on the |ΔS| should be very small, and the large negative lattice expansion at $T_C$ should be the key reason for the very large |ΔS| in LaFe$_{11.4}$Si$_{1.6}$.

As mentioned above, the first-order nature of the phase transition is strengthened with Si content lowering in LaFe$_{13-x}$Si$_x$, and an evolution of the transition from second-order to first-order can take place. For the samples with x ≤ 1.5, thermal and magnetic hysteresis appears inevitably because of the first-order nature of the transition. Details about hysteresis loss can be seen in the following sections. Figure 3b displays the typical Δ*S* as a function of temperature under a field change of 0–5 T for LaFe$_{13-x}$Si$_x$ with different Si content x.[24] The maximal |ΔS| is ~29 J/kgK when x=1.2. However, from a simple analysis it can be concluded that the maximal |ΔS| for LaFe$_{13-x}$Si$_x$ will be ~40 J/kgK. It is easy to see that there should be a one-to-one correspondence between the field-induced magnetization change (Δσ) and ΔS. By comparing the data of different compounds, a Δ*S*-Δσ relation can be obtained, and the utmost entropy change will be the result corresponding to Δσ=1. A remarkable feature is the rapid drop in Δ*S* with x for lower Si content, while slow variation with x for higher Si content as shown in the inset of Figure 3b. The negative lattice expansion near $T_C$ increases rapidly with Si content reducing (see Fig. 3a). It is a signature of the crossover of the magnetic transition from second-order to first-order. These results reveal a fact that large entropy changes occur always accompanied with the first-order phase transition.

### 3.2 Neutron diffraction and Mössbauer studies on LaFe$_{13-x}$Si$_x$

The detailed investigation of magnetic phase transition driven by temperature and magnetic field can give an insight into the mechanism of large magnetic entropy change in these compounds. It is well known that neutron diffraction is a powerful and direct technique to investigate the phase transition, especially for magnetic phase transition. Wang et al.[25] carried out neutron diffraction investigations on LaFe$_{11.4}$Si$_{1.6}$. Rietveld refinements of powder diffraction patterns showed that the occupancy of Fe atoms are ~90.5(±1.7) and ~87.0(±1.8)% for FeI and FeII sites, respectively. Thus Si atoms are almost randomly distributed on these two Fe sites. It is noted that the diffraction profiles at 2K and 300 K can be fitted by one cubic NaZn$_{13}$-type lattice, whereas that at 191 K (very close to the Curie temperature) must be fitted by two cubic NaZn$_{13}$-type lattices with different lattice parameters (Fig. 4). The onset of the ferromagnetic ordering results in a large volume expansion, exerting no influence on the symmetry of the atomic lattice. The volume changes discontinuously and the large volume ferromagnetic phase coexist with a small volume paramagnetic phase at 191 K. The refinement shows that at 191 K, the sample is composed of ~12% of the large volume phase and ~83% of the small one (the rest is ~5% $\alpha$-Fe). The coexistence of two phases implies that the first-order magnetic phase transition and strong interplay between lattice and magnetism take place, which is in agreement with the observations in La(Fe$_{0.88}$Si$_{0.12}$)$_{13}$ from the x-ray diffraction.[26]

It was also found[25] that the lattice parameter is strongly correlated with the Fe moment. With temperature decreasing from 300 to 250 K, the compound displays a normal thermal contraction resulting from the anharmonic vibrations of atoms. Since the Invar effect is caused by the expansion resulting from the spontaneous magnetostriction which cancels the normal thermal contraction, one may infer that the short-range magnetic correlation appears far above the Curie temperature in LaFe$_{11.4}$Si$_{1.6}$. With a further reduction in temperature, the effect of the spontaneous magnetostriction increases and the lattice parameter shows a large jump with a long-range ferromagnetic ordering. Even below the Curie temperature, the contribution of magnetic thermal expansion is still related to the increase of the magnetic correlation as the temperature is lowered.

Although several papers about x-ray diffraction,[26] neutron diffraction,[25] as well Mössbauer studies[27,28] have confirmed two-phase coexistence in LaFe$_{11.44}$Si$_{1.56}$ in a narrow temperature region of $Tc \pm 2$ K, the detailed phase evolutions driven by temperature and magnetic field around $T_C$, especially the individual characteristic feature of these two phases, are not yet well understood. In the case of La(Fe$_{1-x}$Si$_x$)$_{13}$ compounds, a large number of iron atoms provide a unique opportunity to investigate the magnetic phase transition by using $^{57}$Fe Mössbauer spectroscopy. Cheng et al.[29] and Di et al.[30] carried out Mössbauer studies on LaFe$_{13-x}$Si$_x$. Figures 5a and 5b show mössbauer spectra in various applied fields for LaFe$_{11.7}$Si$_{1.3}$ at 190 K and LaFe$_{11.0}$Si$_{2.0}$ at 240 K, respectively. Zero-field Mössbauer spectra collected above $T_C$ show a paramagnetic doublet, owing to the quadrupole splitting. With external field increasing up to 10 kOe, no evidence of a magnetically split sextet is detected in the spectrum of LaFe$_{11.7}$Si$_{1.3}$, but the presence of a sextet is evident in spectrum of LaFe$_{11.0}$Si$_{2.0}$. With external field further increasing over 20 kOe, sharp well-split pairs of sextets are observed, and their areas increase rapidly at the expense of the doublets. Direct evidence of a field-induced magnetic phase transition in LaFe$_{13-x}$Si$_x$ intermetallics with a large magneticaloric effect was provided by $^{57}$Fe Mössbauer spectra in externally applied magnetic fields. Moreover, Mössbauer spectra demonstrate that a magnetic structure collinear to the applied field is abruptly achieved in LaFe$_{11.7}$Si$_{1.3}$ compound once the ferromagnetic state appears, showing a metamagnetic first-order phase transition. In the case of LaFe$_{11.0}$Si$_{2.0}$, the Fe magnetic moments rotate continuously from a random state to the collinear state with applied field increasing, showing that a second-order phase transition is predominant. The different types of phase transformation determine the magnetocaloric effects in response to temperature and field in these two samples.

In the case of Fe-based alloys and intermetallic compounds, $^{57}$Fe hyperfine field is $H_{hf}$ roughly proportional to the Fe magnetic moment $\mu_{Fe}$, and consequently, the temperature dependence of the average hyperfine field for the compounds LaFe$_{13-x}$Si$_x$ with x=1.3, 1.7 and 2.0 can be fitted with Brillouin function (BF) according to the expression

$$H_{hf}(T) = H_{hf}(0) B_{1/2}(h_{hf}(T)/t), \qquad (6)$$

where $B_{1/2}(x) = 2\coth(x) - \coth(x)$ is the Brillouin function, and t=$T/T_c^{BF}$. $T_c^{BF}$ is the temperature of $H_{hf}(T_c^{BF})$=0 obtained from mean field model.

Though the curves can be well fitted at low temperatures, the BF relation fails to fit the $H_{hf}(T)$ near the $T_C$.[20] As in the mean field theory the magnetic phase transition at $T_C$ is presumed to be of the second-order, in the compound LaFe$_{11.7}$Si$_{1.3}$, the significant deviation of the BF relation from the temperature dependence of $H_{hf}(T)$ near $T_C$ suggests that the magnetic phase transition is of the first-order in nature. With Si concentration increasing, the second-order magnetic phase transition is predominant and leads to a smaller magnetic entropy change.

### 3.3 MCE in Co, Mn and magnetic *R* doped LaFe$_{13-x}$Si$_x$

Although the LaFe$_{13-x}$Si$_x$ compounds exhibit a giant MCE, the Δ*S* peak usually appears at low temperatures (< 210 K). For the purpose of practical applications, it is highly desired that the maximal entropy change can take place near the ambient temperature. According to Figure 3b, unfortunately, the MCE weakens rapidly as $T_C$ increases. It is therefore an important issue about how to shift $T_C$ toward high temperatures without significantly affecting Δ*S*.

Hu *et al*.[11,31] found that the best effect could be obtained by replacing Fe with an appropriate amount of Co. An entropy change value of 11.5 J/kgK in LaFe$_{10.98}$Co$_{0.22}$Si$_{1.8}$ at 242 K for a field change of 5 T was observed in 2000.[11] Further study[31] found that the maximal value of Δ*S* in LaFe$_{11.2}$Co$_{0.7}$Si$_{1.1}$ near the Curie temperature $T_C$ of 274 K for a field change of 0–5 T is 20.3 J/kgK, which exceeds that of Gd by a factor of 2 and is nearly as large as those of Gd$_5$Si$_2$Ge$_2$[9] (see the inset of Fig. 6a) and MnFeP$_{0.45}$As$_{0.55}$[14] whereas there was no obvious magnetic hysteresis in the sample, which is highly desired in the sense of practical application. Figure 6a displays the entropy change as a function of temperature for La(Fe$_{1-x}$Co$_x$)$_{11.9}$Si$_{1.1}$. It is very significant that in the La(Fe$_{1-x}$Co$_x$)$_{11.9}$Si$_{1.1}$ (x=0.04, 0.06 and 0.08) the Curie temperature increases from 243 K to 301 K as x increases from 0.04 to 0.08, while |Δ$S$| only slowly decreases from ~23 J/kgK to ~15.6 J/kgK for a field change of 0-5T.[32] The study on the MCE of La$_{0.5}$Pr$_{0.5}$Fe$_{11.5-x}$Co$_x$Si$_{1.5}$ (0≤x≤1.0) was performed by Shen *et al.*[33] The substitution of Co in the La$_{0.5}$Pr$_{0.5}$Fe$_{11.5}$Si$_{1.5}$ causes the order of phase transition at $T_C$ to change from first-order to second-order at x = 0.6. Although the magnetic entropy change decreases with increasing Co concentration, $T_C$ increases from 181 K for x = 0 to 295 K for x = 1.0 and the hysteresis loss at $T_C$ also reduces remarkably from 94.8 J/kg for x = 0 to 1.8 J/kg for x = 0.4 because an increase of Co content can weaken the itinerant electron metamagnetic transition. For the sample of x = 1.0, it is noteworthy that the maximum values of |Δ$S$| at $T_C$ = 295 K for a magnetic field change of 0-2 T and 0-5 T respectively are 6.0 and 11.7 J/kgK, which are 20% higher than those of Gd. The MCE of La$_{1-x}$Pr$_x$Fe$_{10.7}$Co$_{0.8}$Si$_{1.5}$ was also studied.[34] As x grows from 0 to 0.5, the maximal value of entropy change increases from 13.5 to 14.6 J/kgK for a field change of 0–5 T, while $T_C$, which is near room temperature, exhibits only a small change. The effects of substituting Fe by Co on the MCE in LaFe$_{11.7-x}$Co$_x$Si$_{1.3}$,[35] LaFe$_{11.9-x}$Co$_x$Si$_{1.1}$,[35] LaFe$_{11.8-x}$Co$_x$Si$_{1.2}$[36] and LaFe$_{11.4-x}$Co$_x$Si$_{1.6}$[37] were also studied and similar effects to those described above were observed.

The study on the MCE of La$_{0.7}$Pr$_{0.3}$Fe$_{13-x}$Si$_x$ (x=1.5, 1.6, 1.8 and 2.0) exhibited an increase in $T_C$ and a reduction in Δ$S$ due to the substitution of Si for Fe,[38] which is similar to the case of LaFe$_{13-x}$Si$_x$. Although both the Si-doping and the Co-doping drive $T_C$ to high temperatures, the reduction of Δ$S$ is much slower in the latter case. The maximal |Δ$S$| is ~24 J/kgK (Δ$H$=5T) for La$_{0.5}$Pr$_{0.5}$Fe$_{11.5-x}$Co$_x$Si$_{1.5}$ (x=0.3) and ~11 J/kgK (Δ$H$=5T) for La$_{0.7}$Pr$_{0.3}$Fe$_{13-x}$Si$_x$ (x=2.0) while $T_C$ takes nearly the same value of ~218 K. Therefore, reducing the Si content in LaFe$_{13-x}$Si$_x$ and partial replacing Fe by Co is a promising way to get room-temperature giant MCE.

In an attempt to find out a way to depress $T_C$ while effectively retain the large |Δ$S$|, Wang *et al.*[39,40] studied the effect of substituting Fe by Mn, which may have an AFM coupling with adjacent Fe. The Mn content in La(Fe$_{1-x}$Mn$_x$)$_{11.7}$Si$_{1.3}$ is x=0, 0.01, 0.02, and 0.03.[40] The cubic NaZn$_{13}$-type structure keeps unchanged except when the minor α-Fe phase (<5 wt%) for x > 0.02 appears. A decrease in saturation magnetization much larger than that expected due to a simple dilution effect is observed, which is consistent with the anticipated antiparallel arrangements of Fe and Mn. The Curie temperature was found to decrease at a rate of ~174 K for 1% Mn. A large |Δ$S$| was gained in a wide temperature range, though a tendency toward degeneration appears as y increases. It is ~17 J/kgK for $T_C$=130 K and ~25 J/kgK for $T_C$=188 K (Fig. 6b), for a field change of 0-5 T. The temperature span of Δ$S$ increases obviously. It is ~21.5 K for x=0 and ~31.5 K for x=0.03 (Δ$H$=5 T). For La(Fe$_{1-x}$Mn$_x$)$_{11.4}$Si$_{1.6}$,[39] when the content of Mn is high enough (x > 0.06), long-range FM order will be destroyed, and typical spin glass behavior appears.

Effects of Nd substitution on MCE were studied by Anh *et al.*,[41] who declared an increase of $T_C$ and a decrease of MCE in La$_{1-x}$Nd$_x$Fe$_{11.44}$Si$_{1.56}$ (x=0-0.4) with the incorporation of Nd. However, these results are inconsistent with those subsequently obtained by other groups. Fujieda *et al.*[42,43] performed a systematic study on the effect of Ce-doping for the compounds LaFe$_{13-x}$Si$_x$ with x = 1.3, 1.56, and 1.82. It was observed that the substitution of Ce cause $T_C$ to reduce and the entropy and adiabatic temperature changes to increase. Shen et al.[44] studied systematically effects of substituting Fe with R on magnetic properties and MCEs of $La_{1-x}R_xFe_{13-x}Si_x$. It was found that the substitution of R for La in $La_{1-x}R_xFe_{11.5}Si_{1.5}$ (R = Ce, Pr and Nd) leads to a monotonic reduction in $T_C$ due to the lattice contraction as shown in Figure 7a. The $T_C$ exhibits a linear reduction with the decrease of unit-cell volume at a rate of 2990 K/nm$^3$ for R = Ce, 1450 K/nm$^3$ for R = Pr and 800 K/nm$^3$ for R = Nd. Partially replacing La with R causes an enhancement of the field-induced itinerant electron metamagnetic transition, which leads to a remarkable increase in ΔS. The ΔS as shown in Figure 7b for a field change of 0−5 T, varies from 23.7 J/kgK for x=0 to 32.4 J/kgK for $La_{1-x}Pr_xFe_{11.5}Si_{1.5}$ (x=0.5) and to 32.0 J/kgK for $La_{1-x}Nd_xFe_{11.5}Si_{1.5}$ (x=0.3), but keeps at ~24 J/kgK for $La_{1-x}Ce_xFe_{11.5}Si_{1.5}$ (x=0–0.3).

From these results above it is concluded that the substitution of magnetic rare earth R causes a shift of $T_C$ towards low temperatures, and strengthens the first-order nature of the phase transition. Sometimes a second-order phase transition becomes of the first-order after the introduction of R. The MCE is enhanced with the increase of R content.

### 3.4 Interstitial effect in $La(Fe_{1-x}Si_x)_{13}$

For the purpose of practical applications, as mentioned in the previous sections, the giant MCE occurring near the ambient temperature is required. It is therefore highly desired to find out an effective approach to pushing ΔS to high temperatures without reducing its value. In 2002, Chen et al.[45] and Fujita et al.[46] independently found that the incorporation of interstitial hydrogen into $LaFe_{13-x}Si_x$ shifts $T_C$ to high temperatures while a large MCE still appears. For example, the entropy change is as large as 17 J/kgK (ΔH = 5 T) in $LaFe_{11.5}Si_{1.5}H_{1.3}$ even at a temperature of 288 K.[45] The hydrogen concentration was determined by both gas chromatograph and gas fusion analyses. By changing either hydrogen pressure or annealing temperature, under which the sample was processed, Fujieda et al.[47] controlled the concentration of interstitial hydrogen. In contrast, Chen et al.[45,48] tuned the content of hydrogen by carefully regulating the desorption of absorbed hydrogen. The Curie temperature $T_C$ of $LaFe_{13-x}Si_xH_δ$ was found to increase linearly with the increase of hydrogen content δ, while the magnetic transition remained to be of the first-order. This is completely different from the effect of Si- and/or Co-doping, which causes an evolution of magnetic transition from the first order to the second order. In this way, the giant MCE that usually appears at low temperatures can be pushed towards high temperatures. The entropy changes of $LaFe_{11.5}Si_{1.5}H_δ$ (δ=0-1.8)[48] as a function of temperature are shown in Figure 8a. The values of |ΔS| are 24.6 and 20.5 J/kgK (ΔH = 5 T) at 195 and 340 K, respectively. Due to the broadening of magnetic transition caused by hydrogen desorption, the ΔS value in $LaFe_{11.5}Si_{1.5}H_δ$ is somewhat lower in an intermediated hydrogen concentration range. However, the maximum value of |ΔS| for $LaFe_{11.44}Si_{1.56}H_δ$ (δ=0-1.5) keeps at ~23 J/kgK (ΔH = 5 T) as $T_C$ increases from ~195 K to ~330 K.[47]

Wang et al.[49] studied the MCE of hydrides $La(Fe_{1-x}Mn_x)_{11.7}Si_{1.3}H_y$. Although the antiferromagnetic coupling between Fe and Mn causes a decrease of the Curie temperature, $T_C$ still can be tuned around room temperature by controlling the hydrogen absorption and has the values of 287K, 312K and 337K for x=0.01, 0.02 and 0.03, respectively. The effect of hydrogen atoms on $T_C$ is similar to that of the La(Fe,

Si)$_{13}$H$_\delta$ hydrides , for which lattice expansion caused by interstitial atoms depresses the overlap between Fe-3d electrons, thus leading to an increase of $T_C$. The first order phase transition nature weakens after Mn doping, however, the IEM transition remains, which results in a large entropy change (Fig. 8b). The maximal values of |$\Delta S$| are 23.4, 17.7 and 15.9 J/kgK under a magnetic field change from 0 to 5T for x=0.01, 0.02 and 0.03, respectively.

The hydrides are usually chemically unstable above 150 °C, which could be a fatal problem to practical applications. It is therefore necessary to obtain chemically stable interstitial compounds with high $T_C$s and great |$\Delta S$| values. Chen et al.[50,51] studied the effects of interstitial carbon for the LaFe$_{13-x}$Si$_x$C$_\delta$ carbides, which are stable up to the melting point. The LaFe$_{11.6}$Si$_{1.4}$C$_\delta$ ($\delta$=0, 0.2, 0.4, and 0.6) carbides were prepared by the solid-solid phase reaction technique, that is, arc melting Fe-C intermediate alloy together with La, Fe and Si. X-ray diffraction analyses indicate that the cubic NaZn$_{13}$-type structure remains unchanged after the introduction of carbon atoms, but minor α-Fe phase appears when the carbon concentration is $\delta\geq$0.6. The lattice expansions caused by the interstitial carbon atoms are 0.29%, 0.75%, and 0.93% for $\delta$=0.2, 0.4 and 0.6, respectively. Compared with hydrides, carbides have much strong lattice expansion.[52] The Curie temperatures grows from 195 K for $\delta$=0 to 250 K for $\delta$=0.6.

Figure 8c shows the entropy change as a function of temperature for LaFe$_{11.6}$Si$_{1.4}$C$_\delta$.[50] The entropy change is nearly constant when x is below 0.2, but decreases rapidly for x > 0.4. The maximal values of |$\Delta S$|, for a field change of 0–5 T, are 24.2 J/kgK for $\delta$=0.2, 18.8 J/kgK for $\delta$=0.4, and 12.1 J/kgK for x=0.6, respectively. The decrease of |$\Delta S$| for x > 0.4 could be due to impurity phase, which broadens the phase transition. A slightly different carbides LaFe$_{11.5}$Si$_{1.5}$C$_\delta$ was also studied,[51] and similar effects were observed.

Jia et al.[52] studied the effect of interstitial hydrogen on lattice volume of the hydrides LaFe$_{11.5}$Si$_{1.5}$H$_\delta$ ($\delta$=0, 1.2, and 2), based on the Rietveld analyses of powder x-ray diffraction spectra. It was found that the incorporation of interstitial hydrogen causes a lattice expansion while the structural symmetry remains unchanged. Accompanying the lattice expansion, Fe-Fe bond exhibits a concomitant variation. Four of the five Fe-Fe bonds show a tendency towards expansion. The largest elongation occurs for the shortest inter-cluster bond (B$_4$), and the relative change is as large as ~2.37% as $\delta$ increases from 0 to 2. In contrast, the longest Fe-Fe bond (B$_2$) shrinks considerably (−0.53%). The effect of Ce-doping was also studied[52] for comparison. It is fascinating that the increase in Ce content produces essentially the same effect on Fe-Fe bonds as the decrease of hydrogen content, though interstitial atoms occupy different crystallographic sites from rare-earths. A linear increase of Curie temperature with the increase of lattice constant is observed, to be at a rate of ~17790 K/nm for LaFe$_{11.5}$Si$_{1.5}$H$_\delta$/La$_{1-x}$Ce$_x$Fe$_{11.5}$Si$_{1.5}$ and ~10890 K/nm for LaFe$_{11.5}$Si$_{1.5}$C$_\delta$. This is a signature of the strengthening of magnetic coupling. It was found that the change of the shortest Fe-Fe bond dominates the magnetic coupling in the LaFe$_{1s-x}$Si$_x$-based intermetallics. A relation between exchange integral and Fe-Fe distance has been proposed to explain the volume effects observed.

In Table I we give a summary of the magnetic transition temperature $T_C$ and isothermal entropy change |$\Delta S$| for LaFe$_{13-x}$Si$_x$ and related compounds.

## 3.5 Magnetic exchanges in hydrogenised, pressed and magnetic *R*-doped LaFe$_{13-x}$Si$_x$

A remarkable feature of the LaFe$_{13-x}$Si$_x$ compound is the strong dependence of Curie temperature on phase volume. It has been reported that the incorporation of interstitial hydrogen causes a significant increase in $T_C$, while the hydrostatic pressure leads to a reduction in $T_C$. For example, the typical change of the Curie temperature is ~150 K when ~1.6 hydrogen/ f.u. is absorbed and ~ −106 K as the pressure sweeps from 0 to 1 GPa.[55] The most remarkable result is the presence of a universal relation between Curie temperature and phase volume: the former linearly grows with the increase of lattice constant. This result implies the exclusive dependence of the magnetic coupling in LaFe$_{13-x}$Si$_x$ on Fe–Fe distance and no effect of interstitial hydrogen on the electronic structure of the compounds.

The x-ray diffraction measurements for LaFe$_{11.5}$Si$_{1.5}$H$_\delta$ reveal that the introduction of interstitial hydrogen causes a considerable lattice expansion, though the crystal structure remains unchanged. The maximum lattice constant change is ~3.4%. Subsequent magnetic measurements reveal the stabilization of the FM state by interstitial hydrogen, as shown by the increase of $T_C$ from ~190 K to ~356 K.

The direct effect of high pressure or interstitial hydrogen is on phase volume. It is therefore interesting to check the relation between $T_C$ and phase volume. Based on the XRD data collected at the ambient temperature, the lattice constant at $T_C$ (PM phase) varies according to the relation $a(T_C)=a_0-\beta(296-T_C)$ for $T_c \leq 296$ K and $a(T_C)=a_0+\beta(T_C-296)-\Delta a$ for $T_C \geq 296$ K, where $a_0$ is the lattice constant at ~296 K, $\Delta a$ is the lattice expansion accompanying the FM-PM transition, and $\beta/a_0 \approx 8.2 \times 10^{-6}$ K$^{-1}$ is the linear expansivity of LaFe$_{13-x}$Si$_x$. $\Delta a$ can be derived from the rigid shift of the $T_C$-$a_0$ curve along the $a_0$ axis when $T_C$ exceeds 296 K and it is found to be ~0.044 Å, essentially independent of Si content.

To compare with the results of hydrogenation, information on pressure induced volume change is required. The crystal structure of LaFe$_{11.5}$Si$_{1.5}$ was analyzed by synchrotron radiation XRD conducted under the pressures of up to 4.1 GPa. The compressibility obtained is $\kappa=-V^{-1}d\tilde{V}dP\approx 8.639\times 10^{-3}$ GPa$^{-1}$, where $P$ is pressure and $V$ is the volume. The volume under high pressure has the form $V=V_0(1-\kappa P)$, where $V_0$ is the volume under ambient pressure. Based on these results, the $T_C$-$a$ relation under pressures can be obtained (solid circles in Fig. 9). Results of hydrogenation and Ce doping for La$_{1-y}$Ce$_y$Fe$_{11.5}$Si$_{1.5}$ (y=0–0.3) are also presented for comparison (open triangles in Fig. 9). It can be seen that the slopes of the $T_C$-$a$ relations are essentially the same in the cases of hydrogenating and Ce substitution, but considerably large under pressures. This result reveals the exclusive effect of Fe-Fe distance on magnetic coupling.

To improve the magnetic and the magnetocaloric properties, sometimes lanthanum in the materials is partially replaced by magnetic rare earths. In this case the magnetic exchange can also occur between *R* and Fe. It has been reported that both the MCE and the Curie temperature of La$_{1-x}$R$_x$Fe$_{11.5}$Si$_{1.5}$ compounds could be greatly modified by magnetic rare earths as shown in Figure 7. We have found that a magnetic interaction comparable with that among Fe atoms exists between *R* and Fe.[56] It can cause the Curie temperature to enhance up to ~11% when ~30% of the La atoms are replaced by *R*. Further, the *R*–Fe coupling is found to be strongly dependent on the species of rare earths, and monotonically grows as *R* sweeps from Ce to Nd. This could be a consequence of the lanthanide contraction, which causes an enhancement of the intra-atomic magnetic coupling.

The XRD patterns of $La_{0.7}R_{0.3}Fe_{11.5}Si_{1.5}$ were measured. Similar XRD spectra are obtained for all of the samples, suggesting the similar structures of different samples. However, a close view of the XRD patterns shows a continuous, yet considerable, high-angle shift of the Bragg reflection as $R$ sweeps from La to Pr, and Nd. This is a signature of lattice contraction. The maximal and minimal lattice constants are, respectively, ~11.468 Å obtained in $LaFe_{11.5}Si_{1.5}$ and ~11.439 Å obtained in $La_{0.7}Nd_{0.3}Fe_{11.5}Si_{1.5}$.

It was found that the Curie temperatures are ~194, ~173, ~185, and ~188 K, corresponding to La, Ce, Pr, and Nd for $La_{0.7}R_{0.3}Fe_{11.5}Si_{1.5}$, respectively. As expected, obvious decreases of $T_C$ occur after the partial substitution of $R$ for La. A remarkable result is the strong dependence of the doping effects on the $R$ species. Different from the lattice parameter, which displays a monotonic contraction as $R$ goes from La to Ce, Pr, and Nd, $T_C$ decreases along the sequence from La to Nd, Pr, and Ce. The maximal $\Delta a$ appears in $La_{0.7}Nd_{0.3}Fe_{11.5}Si_{1.5}$, whereas the maximal $\Delta T_C$ occurs in $La_{0.7}Ce_{0.3}Fe_{11.5}Si_{1.5}$. This feature remains for other $R$ contents, and a simple analysis shows that $T_C$ decreases with $x$ at a rate of ~20.8 K/atom for $R$=Nd, ~32.3 K/atom for $R$=Pr, and ~85.9 K/atom for $R$=Ce.

Although the doping effect of $R$ varies from sample to sample, the generic tendency is clear: it yields a considerable depression of $T_C$. This result implies the presence of additional factors that affect the Curie temperature by considering the fact that the $R$–$T$ coupling may have a positive contribution to $T_C$. As demonstrated by the data in Figure 10, the incorporation of smaller $R$ atoms leads to significant lattice shrinkage. This, according to our previous work, will cause a depression of the exchange integral between Fe atoms due to the reduction of Fe–Fe distance. From the systematic investigation of the magnetic coupling under high pressure, which yields a lattice contraction without changing sample composition, it has been found that the decrease of phase volume leads to a $T_C$ reduction at a rate of ~3.72 $\tilde{K}$/Å (marked by solid circles in Fig. 10). The Curie temperatures of the $LaFe_{13-x}Si_x$ compounds, with Si content between 1.3 and 1.9, are also presented in Figure 10 for comparison (open circles). A similar $T_C$–$V$ relation to that of the $LaFe_{11.5}Si_{1.5}$ compound under pressure is obtained if only the lattice effects caused by the Si-doping are considered. These results indicate the universality of the $T_C$–$V$ ($V=a^3$) relation for the samples with only the Fe–Fe interaction. It can be clearly seen that the incorporation of $R$ results in a significant change of the $T_C$–$V$ relation (solid squares and triangles in Fig. 10). Although $T_C$ linearly reduces with the decrease of lattice constant, (i.e. the increase of $R$ content), the decrease rate is less rapid than that of $LaFe_{11.5}Si_{1.5}$. This feature becomes increasingly obvious as $R$ goes from Ce to Nd, and a simple calculation gives the $T_C$–$V$ slopes of ~850, ~1470, and ~3520 $\tilde{K}$/nm$^3$, respectively, for the Nd-, Pr-, and Ce-doped compounds. All of these values are smaller than that of $LaFe_{11.5}Si_{1.5}$ (~3720 $\tilde{K}$/nm$^3$) for different extents, and indicate the presence of magnetic coupling between $R$ and Fe.

**3.6 MCE in the vicinity of the first-order phase transition**

For an idealized first-order transition, that is, the magnetization is a step function of temperature, we showed that the Maxwell relation and the Clausius-Clapeyron equation gave similar results.[57] Based on the integrated Maxwell relation, entropy change can be expressed as

$$\Delta S(T) = \int_0^H \left(\frac{\partial M}{\partial T}\right)_{H,P} dH$$

$$= \int_{T_C(0)}^{T_C(H)} \Delta M \delta(T - T_C) \left(\frac{dT_C}{dH}\right)^{-1}_{H,P} dT_C = \frac{\Delta H \Delta M}{\Delta T_C}, \tag{7}$$

where the equalities $(\partial M/\partial T)_{H,P} = -\Delta M \delta(T-T_C)$ and $dT_C/dH = \Delta T_C/\Delta H$ have been used. The right side of Equation (7) is exactly the entropy change predicted by the Clausius-Clapeyron equation. It reveals a constant entropy change in a temperature range between $T_C(0)$ and $T_C(H)$ whereas null otherwise, without the effects from the variation of magnetic order parameter. This work proves the applicability of the Maxwell relation to first-order phase transition.

In reality, a first-order phase transition occurs in a finite temperature range, and two phases may coexist in the transition process. We found[58-61] that in this case the Maxwell relation could yield a spurious $\Delta S$ peak in the vicinity of the Curie temperature $T_C(H=0)$. Besides the phase coexistence, some other important factors may have great influence on the MCE evaluation in the vicinity of first-order phase transition. For example, the magnetic domains and the discrepancy between macroscopic magnetization and magnetic order are usually ignored, which may lead to revaluating the results determined by Maxwell relation in the vicinity of a first-order phase transition.[61]

The origin and physical meaning of spike-like entropy-change curves can be revealed by a comparison between the Maxwell relation and heat capacity methods. As an example, Figure 11a shows the magnetization isotherms of $La_{0.7}Pr_{0.3}Fe_{11.5}Si_{1.5}$ measured in the field ascending process. A stepwise magnetic behavior appears at $T_C$ (critical temperature under zero applied magnetic field), signifying the coexistence of FM and PM phases. The first steep increase of magnetization marks the contribution of the FM phase, while the subsequent stair-like variation signifies the filed-induced FM transition of the PM phase. The corresponding entropy change calculated by the Maxwell relation is shown in Figure 11b ($\Delta H$=5 T). In addition to the flat $\Delta S$ plateau, an extra spike-shaped peak appears at exactly the same temperature where stepwise magnetic behaviors appear. The heat capacities of the two samples were also measured under the fields of 0 and 5 T, and the entropy change indicates the absence of the spike $\Delta S$ peak.

These results show the failure of the Maxwell relation, which cannot give a correct result for the entropy change near $T_C$. Considering the fact that magnetic field affects only the magnetic state of PM phase, which coexists with the FM phase near $T_C$, only the PM phase contributes to thermal effect. With this in mind, a modified equation for calculating $\Delta S$ can be established. Figure 11c is a schematic diagram showing the determination of $\Delta S$ for the system with an idealized stepwise behavior. Denoting the area surrounded by the two $M$-$H$ curves at $T_1$ and $T_2$ as $\Sigma_1 + \Sigma_2$, the Maxwell relation gives $\Delta S = (\Sigma_1 + \Sigma_2)/(T_1 - T_2)$. Considering the fact that the field-induced metamagnetic transition takes place in the PM phase, only $\Sigma_1$ contributes to $\Delta S$. This implies $\Delta S = \Sigma_1/(T_1 - T_2)$.

Stepwise magnetic behaviors widely exist in magnetic materials such as $MnAs_{1-x}Fe_x$[62] and $Gd_5Si_{4-x}Ge_x$.[63] It was also observed in MnAs[64] and $Mn_{1-x}Cu_xAs$[65] under high pressure. It could be a general feature of the first-order phase transition because the finite temperature width of the phase transition. In this case, $\Delta S$ should be handled carefully. It is worthy noting that the applied field drives both magnetic moments and magnetic domains toward the applied field before reaching the saturation magnetization. In fact, magnetization is a measure of magnetic moment in the direction of applied field. Its change in field direction does not inevitably reflect the change in magnetic order of the materials. Only when the magnetization really gives a description of the magnetic order, the Maxwell relation predicts the correct entropy change. The Maxwell relation does not distinguish the FM and PM phases. As a result, when magnetic hysteresis occurs and two phases coexist, the area bounded by two adjacent magnetization isotherms could be large, giving rise to the spike-like $\Delta S$ peak. In contrast, when FM materials are under a saturated field, the applied field can drive the magnetic moments directly, and the domain effect is negligible approximately. Therefore, the Maxwell relation may be applicable when the magnetic moments are manipulated by the applied field freely.

### 3.7 Thermal and magnetic hystereses in La(Fe$_{1-x}$Si$_x$)$_{13}$-based alloys

As mentioned before, significant MCE appears to be usually accompanied with a first-order magnetic transition. A typical feature of the first-order transition is thermal and magnetic hystereses. This phenomenon is especially obvious in R-doped LaFe$_{13-x}$Si$_x$,[44,59] Gd$_5$(Si$_{1-x}$Ge$_x$)$_4$,[63] and MnAs-based[62] compounds. The study by Shen et al.[44] have shown that the $M$–$T$ curve of La$_{1-x}$R$_x$Fe$_{11.5}$Si$_{1.5}$ has a thermal hysteresis, which enhances with the increase of $R$ concentration. The thermal hysteresis is about 1.4 K, 3.2 and 5 K for LaFe$_{11.5}$Si$_{1.5}$, La$_{0.7}$Nd$_{0.3}$Fe$_{11.5}$Si$_{1.5}$ and La$_{0.5}$Pr$_{0.5}$Fe$_{11.5}$Si$_{1.5}$, respectively. This result implies that the substitution of $R$ for La in LaFe$_{11.5}$Si$_{1.5}$ can enhance thermal-induced first-order magnetic transition. Figures 12a, b and c presents the magnetization isotherms of La$_{1-x}$Pr$_x$Fe$_{11.5}$Si$_{1.5}$ (0, 0.2 and 0.4), respectively, obtained for the field descending-ascending cycling.[59] Two features can be clearly seen from these figures. The first one is the enhancement of magnetic hysteresis with the increase of Pr content and the second one is the growth of hysteresis as temperature approaches $T_C$ from above. This result reveals the intensifying of the first-order nature of the phase transition after introducing Pr.

Defining the hysteresis loss as the area encircled by the two magnetization isotherms obtained in the field ascending-descending cycling, Shen *et al.*[44] obtained the temperature dependent hysteresis losses as shown in Figures 12d and e for the La$_{1-x}$R$_x$Fe$_{11.5}$Si$_{1.5}$ compounds (R=Nd or Pr). The hysteresis loss is maximal near $T_C$, reaching a value of ~100 J/kg for R=Pr. With the increase of temperature, it decreases rapidly, and vanishes above ~210 K. This means the weakening of the first-order nature of the phase transition as $T_C$ grows. When temperature is fixed, hysteresis loss increases with the increase of $R$ content. The hysteresis losses are, for instance, ~20 J/kg in LaFe$_{11.5}$Si$_{1.5}$ and ~70 J/kg in La$_{0.7}$Pr$_{0.3}$Fe$_{11.5}$Si$_{1.5}$, under the same temperature of 195 K. Similar effects are observed in the cases of Pr and Nd doping. These results are different from those of Fujieda et al.,[42] who claimed a depression of hysteresis loss after the incorporation of magnetic rare earth Ce.

Magnetic hysteresis can depress the efficiency of magnetic refrigeration. Among other requirements, two basic demands for practical refrigerants are strong MCE and small hysteresis loss. Although the first-order materials have obvious advantages over the second-order ones as far as the entropy change is concerned, they usually exhibit considerable thermal and magnetic hystereses. By partially replacing Ge with Fe, Provenzano et al.[66] depressed the hysteresis loss in the $Gd_5Ge_2Si_2$ compound. Our studies have also shown that appropriate substitution of Co for Fe in $LaFe_{13-x}Si_x$ can cause the weakening of the first-order character of the phase transition, in addition to the high temperature shift of $T_C$, thus a reduction of thermal/magnetic hysteresis.[33,54,67] However, the improvement of the hysteresis behavior always accompanies the weakening of the magnetocaloric property of the materials. It is therefore highly desired to find an approach to depress magnetic hysteresis without considerably spoiling the MCE.

Recently Shen et al.[53] found that the hysteresis can be significantly depressed by introducing interstitial carbon atoms into the compound. Figure 12f displays the hysteresis loss of $La_{0.5}Pr_{0.5}Fe_{11.5}Si_{1.5}C_\delta$. The hysteresis loss decreases from 94.8 J/kgK to 23.1 J/kgK when δ increases from 0 to 0.3. In the meantime, the entropy change, obtained for a field change of 0-5 T, varies from 32.4 J/kgK to 27.6 J/kgK. This result indicates that the introduction of interstitial carbon atoms could be a promising method of depressing hysteresis loss while maintaining MCE. Gao et al.[68] also investigated the entropy change and hysteresis loss in $LaFe_{11.7}(Si_{1-x}Cu_x)_{1.3}$. With Cu content increasing from x = 0 to 0.2, $T_C$ increases from 185K to 200K, metamagnetic behavior becomes weaker, and magnetic entropy change |ΔS| drops off. However, |ΔS| remains a large value, ~20 J/kgK, when x reaches 0.2. Both thermal and magnetic hystereses are reduced by introducing Cu. The maximal hysteresis loss at $T_C$ drops off from 74.1 J/kg to zero when the Cu content x increases from 0 to 0.2.

To get a deep understanding of the effect of magnetic hysteresis, the magnetic isotherms of the $La_{1-x}R_xFe_{11.5}Si_{1.5}$, $La_{0.7}Pr_{0.3}Fe_{13-x}Si_x$, $La_{0.5}Pr_{0.5}Fe_{11.5}Si_{1.5}C_\delta$ and $La_{0.5}Pr_{0.5}Fe_{11.4}Si_{1.6}N_\delta$ intermetallics were further studied. Based on these data the relation between the maximum entropy change and hysteresis loss was established as shown in Figure 13. ΔS and hysteresis loss exhibit a simultaneous change, the former decreases as the latter vanishes. Fortunately, the variation of hysteresis loss is much more rapidly than ΔS, and the latter can be as high as ~20 J/kgK when the former is negligibly small. According to the standard thermodynamics, for the nucleation and the development of the second phase in the background of the first phase, a driving force is required to overcome energy barrier between two phases. These results indicate that the driving force of the phase transition is similar in the $LaFe_{13}$-based intermetallics, regardless of their compositions. The main reason for the reduction of hysteresis loss could be the high temperature shift of $T_C$. The strong thermal fluctuation at a high temperature provides the driving force required by the phase transition.

## 3.8 Direct measurement of MCE for La(Fe,Si)$_{13}$ based compounds

As an alternative characterization of the MCE, the adiabatic temperature change of $La(Fe,Si)_{13}$ was measured by Hu et al.[32,69,70]. Figure 14a displays the temperature-dependent $\Delta T_{ad}$ obtained in both heating and cooling processes for sample $LaFe_{11.7}Si_{1.3}$ ($T_C \approx 188K$). The peak value of $\Delta T_{ad}$ reaches 4 K upon the field changing from 0 to 1.4 T. The field-dependent $\Delta T_{ad}$ collected at different temperatures in the vicinity of $T_C$ is shown in Figure 14b. One can find that $\Delta T_{ad}$ collected above 183.2 K has a nearly linear dependence on applied field in a region of 0.4 T < H < 1.4 T.[69,70] Most curves (except for those at 182.5 and 183.2 K) do not display a saturation behavior. It means that adiabatic temperature change would increase noticeably with field increasing. Based on the linear dependence, $\Delta T_{ad}$ value can reach 5.8K for a 0-2T field change. Similarly, we measured $\Delta T_{ad}$ for La(Fe$_{0.94}$Co$_{0.06}$)$_{11.9}$Si$_{1.1}$ with $T_C$=274K. The observed $\Delta T_{ad}$ reaches 2.4 K upon the field changing from 0 to 1.1 T.[32,70] $\Delta T_{ad}$ also has a nearly linear dependence on applied field at temperatures near $T_C$. In this way, the estimated value of $\Delta T_{ad}$ can be 3.2K for a 0-2T field change.

We also calculated $\Delta T_{ad}$ for La(Fe,Si)$_{13}$-based compounds from the heat capacity measurements. Figure 14c displays the $\Delta T_{ad}$ as a function of temperature for LaFe$_{13-x}$Si$_x$ under different magnetic fields. The maximal value of $\Delta T_{ad}$ as a function of Si content x is shown in Figure 14d. One can find that with the evolution from first-order to second-order transition, $\Delta T_{ad}$ decreases from 11.4 K for x=1.4 to 3.5 K for x=2.2 for a field change of 0−5T. It was also observed that the values of $\Delta T_{ad}$ for LaFe$_{11.7}$Si$_{1.3}$ and LaFe$_{11.1}$Si$_{1.9}$ are 9.4 and 2.6 K for a field change of 0–2T, respectively.

Because of the discrepancy among thermal measurements, the $\Delta T_{ad}$ data reported by different groups are not identical. Fujieda et al.[71] made both indirect and direct $\Delta T_{ad}$ measurements on the same sample, LaFe$_{11.57}$Si$_{1.43}$. The directly measured $\Delta T_{ad}$ value was 6 K at $T_C$ = 188K for a field change of 0–2 T while the indirect $\Delta T_{ad}$ calculated from heat capacity measurement was 7.6 K. The directly measured $\Delta T_{ad}$ for LaFe$_{11.57}$Si$_{1.43}$H$_{1.6}$ was 4 K at $T_C$ = 319 K for a 0–2 T field change, which is probably lower than the indirect value by 2 to 3 K. Despite the discrepance among different experiments, all these data verified the potential application of La(Fe,Si)$_{13}$-based compounds as magnetic refrigerants.

### 3.9 Progress in practical applications

Since Brown proposed to use Gd for the room temperature magnetic cooling in 1976, a number of interesting magnetocaloric materials with tunable Curie temperatures and attractive magnetocaloric properties have been discovered in recent years, such as GdSiGe, La(Fe,Si,Al)$_{13}$, MnFePAs, NiMn(Ga, Sn, In), etc. which have opened the way to improve temperature span and efficiency for a refrigerant device. The synthesis of La(Fe,Si,Al)$_{13}$ is friendly and does not require extremely high-purity and costly raw materials, thus La(Fe,Si,Al)$_{13}$-based materials are considered to have high potential applications near room temperature. Several groups have tested the cooling effect in devices near room temperature. Zimm et al.[72] carried out a preliminary test by using irregular La(Fe$_{0.88}$Si$_{0.12}$)$_{13}$H$_{1.0}$ particles of 250~500 μm in size as refrigerants in a rotary magnetic refrigerator (RMR) and found that cooling capacity of La(Fe$_{0.88}$Si$_{0.12}$)$_{13}$H$_{1.0}$ compares favorably with that of Gd. Fujita et al.[73] tested the hydrogenated La(Fe$_{0.86}$Si$_{0.14}$)$_{13}$ spheres with an average diameter of 500μm in an AMR-type test module, and observed a clear difference in temperature between both the ends of the AMR bed. The temperature span of 16K was achieved in a steady state.

Our prototype test with using various La(Fe,Si,Al)$_{13}$-based particles as refrigerants in an AMR module is being performed. La(Fe,Si,Al)$_{13}$ material is brittle and easily pulverized. Their poor corrosion resistance also restricts their applications. Corrosive characteristics of La(Fe$_{0.94}$Co$_{0.06}$)$_{11.7}$Si$_{1.3}$ with minor α-Fe were investigated by Long et al.[74] It was found that the random pitting corrosion appears first in the phase of 1:13, resulting in products of $La_2O_3$, $\gamma$-Fe(OOH), $Co(OH)_2$ and $H_2SiO_3$. Further studies have revealed that a hybrid inhibitor can prevent the materials from being eroded. Sample tests showed that the best inhibition efficiency was nearly 100% by using the corrosion inhibitor. No corrosion products were found after sample had been immersed in the inhibitor for 7776 h. $La(Fe_{0.92}Co_{0.08})_{11.9}Si_{1.1}$ spheres through a rotating electrode process was also fabricated. The test in an AMR module is under way.

## 4. MCE in La(Fe,Al)$_{13}$-based compounds

$LaFe_{13-x}Al_x$ compounds possess rich magnetic properties compared with $LaFe_{13-x}Si_x$ compounds. For the substitutions of Si for Fe atoms, the stable concentration region is only $1.2 < x < 2.6$ and the obtained pseudobinary compounds exhibit ferromagnetic characteristics.[21,24] By substituting Al for Fe atoms, the concentration region becomes much wider, $1.0 < x < 7.0$, and the stabilized $LaFe_{13-x}Al_x$ compounds exhibit complicated magnetic properties.[19] The systems will be in favor of the $LaFe_4Al_8$ structure if the Al concentration is too large, and a large amount of $\alpha$-Fe will appear if the Al content is too small. $LaFe_{13-x}Al_x$ compounds with $NaZn_{13}$-type structure exhibit three types of magnetic orders with the variation of Al concentration. Mictomagnetic states were found for a high Al concentration from $x = 4.9$ to 7.0, originating from a competition between antiferromagnetic Fe-Al-Fe superexchange and ferromagnetic Fe-Fe direct exchange. For the Al concentration ranging from $x=1.8$ to 4.9, the system manifests soft ferromagnetic properties. At the minimum permitted Al concentrations from $x=1.0$ to 1.8, they show weak antiferromagnetic coupling, which can be overcome even by applying a small field of a few Testa and cause a spin-flip transition to ferromagnetic state. Our studies have revealed that a small doping of Co can make the antiferromagnetic coupling collapse, resulting in a ferromagnetic state.

### 4.1 Room temperature MCE in Co-doped La(Fe,Al)$_{13}$

In 2000, Hu et al.[11,75] firstly studied magnetic entropy change in Co-doped La(Fe,Al)$_{13}$. $La(Fe_{0.98}Co_{0.02})_{11.7}Al_{1.3}$ and $LaFe_{11.12}Co_{0.71}Al_{1.17}$ exhibit ferromagnetic behaviours with a second-order magnetic transition at $T_C\sim198K$ and $\sim279K$, respectively. The magnetic entropy change is about 5.9 and 10.6 J/kgK for $La(Fe_{0.98}Co_{0.02})_{11.7}Al_{1.3}$, and 4.6 and 9.1 J/kgK for $LaFe_{11.12}Co_{0.71}Al_{1.17}$ under field changes of 0–2 T and 0–5 T, respectively. Our experiments have confirmed the antiferromagnetic nature of $LaFe_{11.7}Al_{1.3}$ and $LaFe_{11.83}Al_{1.17}$, and found that Co doping can convert the antiferromagnetic coupling to a ferromagnetic one. $T_C$ shifts toward higher temperatures with Co content increasing.[76] Figure 15 displays magnetic entropy change of $La(Fe_{1-x}Co_x)_{11.83}Al_{1.17}$ ($x=0.06$ and 0.08). The $\Delta S$ of $La(Fe_{1-x}Co_x)_{11.83}Al_{1.17}$ has nearly the same magnitude as that of Gd near room temperature. The calculated $\Delta S$ in the molecular field approximation is also shown in Figure 15. The theoretical result is in qualitative agreement with the experimental one. Since the highest magnetocaloric effect involving a second order magnetic transition near room temperature is produced by Gd, and most intermetallic compounds which are ordered magnetically near or above room temperature show significantly lower $|\Delta S|$ than Gd,[7] obviously, these results are very attractive.

The high magnetization of Co-doped La(Fe,Al)$_{13}$ is considered to be responsible for the large $|\Delta S|$.[76]

From *M-H* curves measured at 5 K, the values of 2.0 and 2.1$\mu_B$/Fe(Co) were determined for La(Fe$_{1-x}$Co$_x$)$_{11.83}$Al$_{1.17}$ (*x*=0.06 and 0.08), respectively. Usually, a small substitution of Co can shift $T_C$ toward high temperatures without affecting the saturation magnetization considerably. As a result, |ΔS| remains nearly unchanged upon increasing the substitution of Co for Fe.

**4.2 Nearly constant magnetic entropy change in La(Fe, Al)$_{13}$**

An ideal magnetic refrigerant suitable for use in an Ericsson-type refrigerator should have a constant (or almost constant) magnetic entropy change through the thermodynamical cycle range.[77] A good choice for a suitable Ericsson-cycle refrigerant would be a single material with an appropriate |ΔS| profile. Typical materials with such properties are those of the series (Gd, Er)NiAl,[78] in which the suitable working temperature range is from ~10 to ~80K. However, at relatively high temperatures, rare materials were reported to show a table-like Δ*S*. We investigated magnetic entropy change around phase boundary in LaFe$_{13-x}$Al$_x$ compounds. A table-like Δ*S* from ~140K to ~210 K involving two successive transitions was found in a LaFeAl sample at phase boundary.[79]

Hu *et al*. tuned Al content from x = 1.82 to x = 1.43 in LaFe$_{13-x}$Al$_x$ and a gradual change from ferromagnetic (FM) to weak antiferromagnetic (AFM) state was observed. A completely FM ground state at *x*=1.82 is followed by the emergence of AFM coupling at *x*=1.69 and 1.56, in which two spaced transitions appear, one at $T_0$ from FM to AFM and the other at $T_N$ from AFM to paramagnetic state. The transition nature at $T_0$ and $T_N$ are of first-order and second-order, respectively. Continuously reducing Al to x=1.43 results in a completely AFM ground state.[80] X-ray diffraction measurements at different temperatures were performed to monitor the change of crystal structure. We found that the samples remain cubic NaZn$_{13}$-type structure when the magnetic state changes with temperature, but the cell parameter changes dramatically at the first-order transition point $T_0$.[79]

From the magnetic entropy change Δ*S* as functions of temperature and magnetic field for LaFe$_{13-x}$Al$_x$ (x=1.82, 1.69, 1.56, and 1.43) compounds, it was found that with the emergence and enhancement of AF coupling, the ΔS profile evolves from a single-peak shape at x=1.82 to a nearly constant-peak shape at *x*=1.69 and 1.56, and then to a two-peak shape at 1.43. The nearly temperature independent |ΔS| over a wide temperature range (an about 70 K span from ~140 to 210 K) in the sample with *x*=1.69 is favorable for application in an Ericsson-type Refrigerator working in a corresponding temperature range.

**4.3 Interstitial Effect in La(Fe,Al)$_{13}$**

Wang et al.[81] investigated interstitial effect on magnetic properties and magnetic entropy change in La(Fe, Al)$_{13}$ alloys. Carbonization brings about an obvious increase in lattice parameter, thus an antiferromagnetic to ferromagnetic transition. In LaFe$_{11.5}$Al$_{1.5}$, a considerable increase of Curie temperature from 191 to 262 K was observed with carbon concentration increasing from 0.1 to 0.5, however, only a slight increase in saturation magnetization is accompanied. The magnetic transition is of second-order in nature and thus the magnetization is fully reversible on temperature and magnetic field. One can find that all the LaFe$_{11.5}$Al$_{1.5}$ carbides exhibit a considerable magnetic entropy change, comparable with that in Gd around the Curie temperature. Thus, one can get a large reversible magnetic entropy change over a wide temperature range by controlling the carbon concentration.

## 5. Magnetic and magnetocaloric properties in Mn-based Heusler alloys

Mn-based Heusler alloys are well-known for their shape-memory effect, superelasticity, and magnetic-field-induced strain. In stoichiometric $Ni_2MnGa$ alloys, the nearest neighboring distance between Mn atoms is around 0.4 nm. Ruderman-Kittel-Kaeya-Yo (RKKY) exchange through conductive electrons leads to a ferromagnetic ordering. The magnetic moment is mainly confined to Mn atoms, ~4.0 $\mu_B$, while Ni has a rather small moment. Ni–Mn–Ga undergoes a martensitic-austenitic transition. Although both the martensite and austenite phases are usually ferromagnetic, their magnetic behaviors are significantly different. The martensitic phase is harder to be magnetically saturated because of its large magnetocrystalline anisotropy. The simultaneous changes in structure and magnetic property at the phase transition yield significant entropy changes.

In 2000, Hu *et al.* firstly reported on entropy change $\Delta S$ associated with the structure transition in a polycrystalline $Ni_{51.5}Mn_{22.7}Ga_{25.8}$ (Fig. 16a).[82] The martensitic-austenitic transition, which is of first order in nature with a thermal hysteresis about 10K, takes place at ~197K. A positive entropy change, 4.1J/kgK for a field change of 0–0.9 T, appears, accompanying the field-induced changes in magnetization and magnetic anisotropy. A positive-to-negative crossover of the entropy change[83] and subsequent growth in magnitude were further observed as magnetic field increases, and a negative $\Delta S$ of ~ −18 J/kgK (300K) for the field change of 0–5 T was obtained in single crystal $Ni_{52.6}Mn_{23.1}Ga_{24.3}$ (Fig. 16b).[84] For the $Ni_{52.6}Mn_{23.1}Ga_{24.3}$ single crystal, the thermal hysteresis around martensitic transition is about 6 K, and the magnetic hysteresis is negligible.

Since the first report of large entropy change in $Ni_{51.5}Mn_{22.7}Ga_{25.8}$,[82] numerous investigations on the magnetic properties and magnetocaloric effect (MCE) in various ferromagnetic shape memory Heusler alloys (FSMAs) have been carried out.[16] Typical reports are about the great |$\Delta S$| obtained in a polycrystalline $Ni_2Mn_{0.75}Cu_{0.25}Ga$[85] and a single crystal $Ni_{55.4}Mn_{20.0}Ga_{24.6}$,[86] in which the structural and the magnetic transitions are tuned to coincide with each other. However, the large entropy change usually appears in a narrow temperature range, for example 1−5 K. As is well known, a real magnetic refrigerator requires not only a large MCE but also a wide temperature span of the MCE. Although the |$\Delta S$| in these conventional Heusler alloys can be very large, the narrow temperature span of the $\Delta S$ may restrict their applications.

A recent discovery of metamagnetic shape memory alloys (MSMAs) has arouse the intensive interest because of huge shape memory effect and different mechanism from that of traditional alloys.[87] In these Ga-free Ni-Mn-Z Heusler alloys (where Z can be an element of group III or group IV, such as In, Sn or Sb), an excess of Mn causes a fundamental change of magnetism for parent and product phases. A strong change of magnetization across the martensitic transformation results in a large Zeeman energy $\mu_0\Delta M \cdot H$. The enhanced Zeeman energy drives the structural transformation and causes a field-induced metamagnetic behavior, which is responsible for the huge shape memory effect. The simultaneous changes in structure and magnetism, induced by magnetic field, should be accompanied by a large MCE. Several groups studied magnetic properties and MCE, and inverse MCE with a relative wide temperature span has been observed.[88-92]

The compositions of $Ni_{50}Mn_{34}In_{16}$ belong to the so-called MSMAs, which is the only one that exhibits a field-induced transition in $Ni_{50}Mn_{50-y}In_y$.[93] Several groups investigated its shape memory effect and magnetocaloric effect. The reported ΔS with a considerable large temperature span reaches 12J/kgK under a magnetic field of 5T,[92] which is larger than that of Gd. However, the large ΔS takes place around 180 K, which is still far from room temperature. Furthermore, a large hysteresis is accompanied even for the metamagnetic shape memory alloys. The reported thermal hysteresis can be as large as ~20 K for Ni-Mn-Sn[89] and ~10K for Ni-Co-Mn-In alloys.[87] For $Ni_{50}Mn_{34}In_{16}$, the thermal hysteresis even reaches 20 K, and more seriously it becomes further wider with external field increasing.[92] Our recent studies revealed that a little more increase of Ni content not only increases $T_m$ and but also significantly enhances the magnetic entropy change. More importantly, it can remarkably improve the thermal hysteresis.[94]

The zero-field-cooled (ZFC) and field-cooled (FC) magnetizations were measured under 0.05 T, 1 T, and 5 T for $Ni_{51}Mn_{49-x}In_x$ (x = 15.6, 16.0, and 16.2).[94] One can find that all alloys show a very small thermal hysteresis, < 2 K, around martensitic transition. More importantly, an increase in magnetic filed does not enlarge the hysteresis for all samples. The frictions from domain rearrangements and phase boundary motions are considered to be a main factor affecting the hysteresis gap.[95,96] The gap of thermal hysteresis may characterize the strength of frictions during the transformation. In these systems, the small hysteresis indicates that the friction to resist the transformation is small. Anyway, the small thermal hysteresis is an aspiration of engineers to apply MCE materials to a refrigerator. These features guarantee that the magnetocaloric effect is nearly reversible on temperature even a high magnetic field is applied.

Our study showed that the entropy change ΔS of $Ni_{51}Mn_{49-x}In_x$ is positive, peaks at $T_m$ and gradually broadens to lower temperature, which is a result of the field induced metamagnetic transition from martensitic to austenitic state at temperatures below $T_m$. The maximal values of ΔS reach 33, 20, and 19 J/kgK at 308 K, 262 K, and 253 K for compositions x = 15.6, 16.0, and 16.2, respectively. In comparison with ΔS (12 J/kgK, at 188 K) observed in $Ni_{50}Mn_{34}In_{16}$ alloys, not only the $T_m$, at which ΔS peaks, goes much nearer to room temperature but also the size of ΔS is remarkably enhanced. ΔS span could reach ~20 K under a field of 5T. Such a large temperature span should be attractive compared with the traditional Heusler alloys.

The ΔS shows a table-like peak under 5T for $Ni_{51}Mn_{49-x}In_x$. The flat plateau of ΔS should reflect the intrinsic nature of magnetocaloric effect. In some first-order systems, such as LaFeSi, ΔS peak usually exhibits a peculiar shape, an extremely high spike followed by a flat plateau. Detailed studies[58] suggested that the extremely high peak dose not reflect the intrinsic entropy change but a spurious signal. However, careful investigations based on specific heat measurements verified that the flat plateau does reflect the intrinsic nature of ΔS. Similar to the case of the first-order systems La-Fe-Si,[58] the broad plateau of ΔS should reflect the intrinsic nature of magnetocaloric effect. As is well known, a plateau-like ΔS is specially desired for Ericsson-type refrigerators.

We also investigated magnetic properties and entropy change in Co-doped NiMnSn alloys, and found that the incorporation of Co enhances ferromagnetic exchange for parent phases, while the magnetic exchange of martensitic phase keeps nearly unchanged. An external magnetic field can shift $T_m$ to a lower temperature at a rate of 4.4 K/T in $Ni_{43}Mn_{43}Co_3Sn_{11}$ and a field-induced structural transition takes place. Associated with the metamagnetic behaviors, a large positive entropy change, ~33 J/kgK ($\Delta H$=5 T, at 188 K), is observed. The $\Delta S$ also displays a table-like peak under 5 T.

## 6. MCE in other materials

Compared with the first-order materials, the second-order ones can have comparable or even larger refrigerant capacity (*RC*) though they sometimes exhibit relatively low $\Delta S$. Moreover, the absences of magnetic and thermal hystereses are also promising features of the materials of this kind. As mentioned above, the hysteresis loss, which makes magnetic refrigeration less efficient, usually happened accompanying a first-order transition. It is therefore of significance to search for efficient magnetic refrigerants with the second-order characters.

### 6.1 MCE in $R_6Co_{1.67}Si_3$

In the previous studies, a family of ternary silicides $R_6Ni_2Si_3$ with R = La, Ce, Pr and Nd was discovered.[97] Recently, a ferromagnetic silicide $Gd_6Co_{1.67}Si_3$ derived from the $Ce_6Ni_2Si_3$-type structure was reported.[98] The compound exhibits a high saturation magnetization and a reversible second-order magnetic transition at a temperature of 294 K. Thus, large values of $|\Delta S|$ and *RC* of $R_6Co_2Si_3$ compounds around room temperature could be expected. Shen *et al*.[99-101] studied the magnetic properties and MCEs of the $R_6Co_{1.67}Si_3$ compounds with R = Pr, Gd and Tb. The MCE of $R_6Co_{1.67}Si_3$ (R = Gd and Tb) and $Gd_6M_{5/3}Si_3$ (M = Co and Ni) were also studied by Jammalamadaka *et al*.[102] and Gaudin *et al*.,[103] respectively.

$R_6Co_{1.67}Si_3$ (R = Pr, Gd and Tb) have a single phase with a hexagonal $Ce_6Ni_2Si_3$-type structure (space group P6$_3$/m). The $T_C$s are determined to be 48, 298 and 186 K, respectively. The $T_C$ of $Gd_6Co_2Si_3$ is nearly as large as that of Gd. Figure 17a shows the magnetization isotherms of $R_6Co_2Si_3$ (R = Pr, Gd and Tb) around the Curie temperature.[99-101] It is evident that each isotherm near $T_C$ shows a reversible behaviour between the increasing field and decreasing field. Moreover, neither inflection nor negative slope in the Arrott plot of $R_6Co_{1.67}Si_3$ is observed as a signature of metamagnetic transition above the $T_C$, indicating a characteristic of second-order magnetic transition. The $\Delta S$ as a function of temperature for the $R_6Co_{1.67}Si_3$ compounds with *R* = Pr, Gd and Tb is shown in Figure 17b.[99-101] It is found that both the height and the width of $\Delta S$ peak depend on the applied field, increasing obviously with the increase of applied filed. There is no observed a visible change in peak temperature of $\Delta S$. The $\Delta S$–*T* curve shows a "λ"-type as displayed in typical second-order magnetocaloric materials. For the $R_6Co_{1.67}Si_3$ (R = Pr, Gd and Tb), the maximal values of $|\Delta S|$ are 6.9, 5.2 and 7.0 J/kgK, respectively, for field changing from 0 to 5T.

In general, the refrigeration capacity (RC) is an important characteristic of the magnetocaloric materials, providing an accepted criterion to evaluate the refrigeration efficiency which is of specially importance in practical application. The *RC* value, obtained by integrating numerically the area under the $\Delta S$-*T* curve using the temperature at half maximum of the $\Delta S$ peak as the integration limits,[104] is 440 J/kg for $Gd_6Co_{1.67}Si_3$ for a field change of 0–5 T, much larger than those of some magnetocaloric materials for a field change of 0–5 T, such as $Gd_5Ge_{1.9}Si_2Fe_{0.1}$ (~355 J/kg at 305 K),[66] $Gd_5Ge_{1.9}Si_2$ (~235 J/kg at 270 K),[66] and $Gd_5Ge_{1.8}Si_{1.8}Sn_{0.4}$ ribbons prepared at 15–45 m/s (~305−335 J/kg at ~260 K).[105] For the *RC* value, it is necessary to take into account the hysteresis loss. However the study on the isothermal field dependence of magnetization for $Gd_6Co_{1.67}Si_3$ reveals no hysteresis loss. It is very important to $Gd_6Co_{1.67}Si_3$ that the large |$\Delta S$| and the enhanced *RC* are observed to occur around 298 K, thereby allowing room-temperature magnetic refrigeration. This result is of practical importance, because the $Gd_6Co_{1.67}Si_3$ can be a good working material for magnetic refrigeration at the ambient temperature.

**6.2 MCE in $CdCr_2S_4$**

$AB_2X_4$-type sulfospinels have attracted much attention due to their colossal magnetocapacity effects[106] and large magnetoresistance effects.[107] Many of sulfospinels, e.g. $(Cd,Hg)Cr_2(S,Se)_4$, have ferromagnetic spin configuration and large spontaneous magnetization.[108] $CdCr_2S_4$ is a member of the chalcogenide $ACr_2S_4$ spinels with ferromagnetically coupled $Cr^{3+}$ spins (*S*=3/2). Recently, Yan et al.[109] studied the magnetocaloric effects of $CdCr_2S_4$. A polycrystalline $CdCr_2S_4$ sample was fabricated by using the solid-state reaction method. The sample is a normal spinel structure of space group *Fd*3*m* with $Cr^{3+}$ octahedrally and $Cd^{2+}$ tetrahedrally surrounded by sulfur ions, and its lattice parameter and Curie temperature are 1.0243(4) nm and = 87 K, respectively. The saturation moment is about 5.96 $\mu_B$ per formula unit, in agreement with ferromagnetically ordered $Cr^{3+}$ spins due to the superexchange interaction between Cr–S–Cr atoms.[110] Magnetic entropy change versus temperature is shown in Figure 18(a). Near the Curie temperature, the maximal entropy change is 3.9 and 7.0 J/kg K for the field changes of 2 and 5 T, respectively. The Arrott plot of $CdCr_2S_4$ shows a characteristic of second-order magnetic transition. This large magnetic entropy change can be attributed to a sharp drop in magnetization with temperature increasing near Curie temperature. Yan et al also performed the measurements of the heat capacity in the fields of *H*=0, 2, and 5 T. An applied field broadens the peak and rounds it off in high fields, which further indicates a second-order phase transition.[9] The isothermal magnetic entropy change $\Delta S_{heat}$ calculated from the heat capacity data exhibits a similar behavior to $\Delta S$. The adiabatic temperature change $\Delta T_{ad}$ is presented in Fig. 18(b). The maximal values of $\Delta T_{ad}$ are about 1.5 and 2.6 K for magnetic field changes of 2 and 5 T, respectively. Shen et al.[111] further studied the magnetocaloric effects in spinels (Cd, M)$Cr_2S_4$ with M = Cu or Fe. It is found that the partial replacement of Cd by Cu can exert a little influence on the magnetic coupling, and only a small shift of $T_C$ from 86 K to 88 K was observed. In contrast, a significant increase of $T_C$ from 86 K to 119 K was observed, which stems from the substitution of Fe for Cd. The maximal values of magnetic entropy change $\Delta S$ were found to be 5.1 and 5.4 J/kgK for $Cd_{0.8}Cu_{0.2}Cr_2S_4$ and $Cd_{0.7}Fe_{0.3}Cr_2S_4$ for a field change from 0 to 5T, respectively.

### 6.3 MCE in amorphous alloys

Amorphous magnetic materials, in spite of their relatively small magnetic entropy change compared with that of crystalline materials, usually have a large refrigerant capacity. Recently, magnetic entropy change and refrigerant capacity (RC) of Gd-based amorphous $Gd_{71}Fe_3Al_{26}$ and $Gd_{65}Fe_{20}Al_{15}$ alloys were investigated by Dong et al.[112] The values of $T_C$ are 114 K for $Gd_{71}Fe_3Al_{26}$ and 180 K for $Gd_{65}Fe_{20}Al_{15}$, respectively, which can be easily tunable by adjusting the composition. Furthermore, scarcely any thermal hysteresis can be observed in the vicinity of $T_C$. The maximal value of | magnetic entropy change $\Delta S$/ (7.4 J/kgK for $Gd_{71}Fe_3Al_{26}$ and 5.8 J/kgK for $Gd_{65}Fe_{20}Al_{15}$, $\Delta H$=0−5T) is not very large, however, the values of RC reach 750 J/kg and 726 J/kg for $Gd_{71}Fe_3Al_{26}$ and $Gd_{65}Fe_{20}Al_{15}$, respectively, which are much larger than those of other magnetocaloric materials ever reported. Such a high *RC* is due to the glassy structure that extends the large MCE into a broad temperature range.

Wang *et al.*[113] studied the magnetic properties and MCEs of amorphous $Ce_2Fe_{23-x}Mn_xB_3$ ($1 \leq x \leq 6$) alloys. It was found that the magnetic state is sensitive to the Mn content due to the competition between the Fe-Fe ferromagnetic coupling and the Fe–Mn antiferromagnetic coupling. A spin glass behavior at low temperatures was observed in the samples of $x \geq 4$. Typical ferromagnetism appears for the samples with $x \leq 3$, and their $T_C$ almost linearly decreases from 336 to 226 K as x increases from 1 to 3. The magnetization has a sharp drop around $T_C$ without thermal hysteresis, suggesting a second order phase transition resulting from their amorphous nature. In spite of the relatively small |ΔS|, the value of *RC* for amorphous $Ce_2Fe_{22}MnB_3$ alloy is found to be ~225 J/kg ($\Delta H$=0–5 T), which is comparable with that of some good crystalline materials with $T_C$ around room temperature.

### 7. Summary and outlook

Investigations on magnetocaloric effect (MCE) are of great importance for not only fundamental problems but also technological applications. Over the past decade, we have investigated the MCE and relevant physics in several kinds of materials, including La(Fe, *M*)$_{13}$-based compounds with *M*=Si and Al, NiMn-based Heulser alloys, $Ce_6Ni_2Si_3$-type $R_6Co_{1.67}Si_3$ compounds, $AB_2X_4$-type sulfospinels $CdCr_2S_4$, etc. Among these materials, the La(Fe, *M*)$_{13}$-based compounds have received most attention. We have found a large entropy change (|ΔS| > 19 J/kgK, at $T_C$ < 210 K) for a field change of 0−5 T in La(Fe, Si)$_{13}$ with a low Si concentration, which is associated with negative lattice expansion and metamagnetic transition behaviour. Partially replacing La with magnetic *R* atoms in La(Fe, Si)$_{13}$ leads to a remarkable increase in entropy change, a reduction in $T_C$ and an increase in magnetic hysteresis. A strong MCE and zero hysteresis loss are obtained near room temperature in Co-doped La(Fe, Si)$_{13}$ alloys. By introducing interstitial hydrogen atoms, we have found that the large MCE can remain at room temperature. The maximal value of |ΔS| for $LaFe_{11.5}Si_{1.5}H_\delta$ attains to 20.5 J/kgK at 340 K for a field change of 0−5 T, which exceeds that of Gd by a factor of 2. Introducing interstitial carbon atoms is found to be a promising method of depressing hysteresis loss while the large MCE is kept unchanged. To understand the nature of the large MCE, the detailes of phase volume and magnetic exchanges are studied in hydrogenised, pressed and magnetic *R*-doped $LaFe_{13-x}Si_x$ alloys**.** The most remarkable result we have obtained is the presence of a universal relation between Curie temperature and phase volume: the Curie temperature linearly goes up with the increase of lattice constant. This result implies the exclusive dependence of the magnetic coupling in LaFe$_{13-x}$Si$_x$ on Fe–Fe distance.

La(Fe, $M$)$_{13}$-based materials have attracted worldwide attention in recent years. Several hundred scientific papers dealing with these materials have been published since the advent of the first report on their large entropy change in 2000. The low cost, easy and friendly preparation, and large magnetocaloric effect near room temperature make La(Fe, $M$)$_{13}$-based compounds more attractive as candidates for magnetic refrigerants, especially the potential application near room temperature. La(Fe, $M$)$_{13}$-based compounds may be a good candidate that can replace Gd metal as a room temperature refrigerant. Up to now, over 30 related patents have been published all over the world. Several groups have tested the cooling effect of La(Fe, $M$)$_{13}$-based materials in principle-of-proof experiments of refrigerators. The cooling capacity near room temperature has been verified. All these efforts and achievements give us fresh hope that the magnetic refrigerators based on La(Fe, $M$)$_{13}$-based refrigerants may come ture in industry and daily life in the near future.

**Acknowledgements**

This work was supported by the National Basic Research Program of China, the National Natural Science Foundation of China and the Knowledge Innovation Project of the Chinese Academy of Sciences.

Table 1 Magnetic transition temperature $T_C$ and isothermal entropy change $|\Delta S|$ for $LaFe_{13-x}Si_x$ and related compounds

| compounds | $T_C$ (K) | $-\Delta S$ (J/kgK) 0–2 T | $-\Delta S$ (J/kgK) 0–5 T | Refs. |
|---|---|---|---|---|
| $LaFe_{11.83}Si_{1.17}$ | 175 | 21.2 | 27.8 | This work |
| $LaFe_{11.8}Si_{1.2}$ | | 25.4 | 29.2 | This work |
| $LaFe_{11.7}Si_{1.3}$ | 183 | 22.9 | 26.0 | This work |
| $LaFe_{11.6}Si_{1.4}$ | 188 | 20.8 | 24.7 | This work |
| $LaFe_{11.5}Si_{1.5}$ | 194 | 20.8 | 24.8 | This work |
| $LaFe_{11.5}Si_{1.5}$ | 194 | 21.0 | 23.7 | [44] |
| $LaFe_{11.5}Si_{1.5}$ | 195 | 21.9 | 24.6 | [48] |
| $LaFe_{11.4}Si_{1.6}$ | 199 | 14.2 | 18.7 | This work |
| $LaFe_{11.4}Si_{1.6}$ | 209 | 14.3 | 19.3 | [11] |
| $LaFe_{11.4}Si_{1.6}$ | 208 | 14.3 | 19.4 | [12] |
| $LaFe_{11.3}Si_{1.7}$ | 206 | 11.9 | 17.6 | This work |
| $LaFe_{11.2}Si_{1.8}$ | 210 | 7.5 | 13.0 | This work |
| $LaFe_{11.0}Si_{2.0}$ | 221 | 4.0 | 7.9 | This work |
| $LaFe_{10.6}Si_{2.4}$ | | 2.8 | | [24] |
| $LaFe_{10.4}Si_{2.6}$ | | 2.3 | | [14] |
| $La_{0.9}Pr_{0.1}Fe_{11.5}Si_{1.5}$ | 191 | ~24 | 26.1 | [44] |
| $La_{0.8}Pr_{0.2}Fe_{11.5}Si_{1.5}$ | 188 | ~26 | 28.3 | [44] |
| $La_{0.7}Pr_{0.3}Fe_{11.5}Si_{1.5}$ | 185 | ~28 | 30.5 | [44] |
| $La_{0.6}Pr_{0.4}Fe_{11.5}Si_{1.5}$ | 182 | ~29 | 31.5 | [44] |
| $La_{0.5}Pr_{0.5}Fe_{11.5}Si_{1.5}$ | 181 | ~30 | 32.4 | [44] |
| $La_{0.9}Nd_{0.1}Fe_{11.5}Si_{1.5}$ | 192 | ~23 | 25.9 | [44] |
| $La_{0.8}Nd_{0.2}Fe_{11.5}Si_{1.5}$ | 190 | ~24 | 27.1 | [44] |
| $La_{0.7}Nd_{0.3}Fe_{11.5}Si_{1.5}$ | 188 | ~29 | 32.0 | [44] |
| $La_{0.7}Ce_{0.3}Fe_{11.5}Si_{1.5}$ | 173 | | 23.8 | [44] |
| $La_{0.7}Pr_{0.3}Fe_{11.4}Si_{1.6}$ | 190 | 25.4 | 28.2 | [38] |
| $La_{0.7}Pr_{0.3}Fe_{11.2}Si_{1.8}$ | 204 | 14.4 | 19.4 | [38] |
| $La_{0.7}Pr_{0.3}Fe_{11.0}Si_{2.0}$ | 218 | 6.2 | 11.4 | [38] |
| $LaFe_{11.5}Si_{1.5}H_{0.3}$ | 224 | | 17.4 | [48] |
| $LaFe_{11.5}Si_{1.5}H_{0.6}$ | 257 | | 17.8 | [48] |
| $LaFe_{11.5}Si_{1.5}H_{0.9}$ | 272 | | 16.9 | [48] |
| $LaFe_{11.5}Si_{1.5}H_{1.3}$ | 288 | 8.4 | 17.0 | [45,48] |
| $LaFe_{11.5}Si_{1.5}H_{1.5}$ | 312 | | 16.8 | [48] |
| $LaFe_{11.5}Si_{1.5}H_{1.8}$ | 341 | | 20.5 | [48] |
| $La(Fe_{0.99}Mn_{0.01})_{11.7}Si_{1.3}H_\delta$ | 336 | 16.0 | 23.4 | [49] |
| $La(Fe_{0.98}Mn_{0.02})_{11.7}Si_{1.3}H_\delta$ | 312 | 13.0 | 17.7 | [49] |
| $La(Fe_{0.97}Mn_{0.03})_{11.7}Si_{1.3}H_\delta$ | 287 | 11.0 | 15.9 | [49] |
| $LaFe_{11.5}Si_{1.5}C_{0.2}$ | 225 | 18.0 | 22.8 | [51] |
| LaFe$_{11.5}$Si$_{1.5}$C$_{0.5}$ | 241 | 7.4 | 12.7 | [51] |
| La$_{0.5}$Pr$_{0.5}$Fe$_{11.5}$Si$_{1.5}$C$_{0.3}$ | 211 | 25.2 | 27.6 | [53] |
| La(Fe$_{0.96}$Co$_{0.04}$)$_{11.9}$Si$_{1.1}$ | 243 | 16.4 | 23.0 | [32] |
| La(Fe$_{0.94}$Co$_{0.06}$)$_{11.9}$Si$_{1.1}$ | 274 | 12.2 | 19.7 | [32] |
| La(Fe$_{0.92}$Co$_{0.08}$)$_{11.9}$Si$_{1.1}$ | 301 | 8.7 | 15.6 | [32] |
| LaFe$_{11.2}$Co$_{0.7}$Si$_{1.1}$ | 274 | | 20.3 | [31] |
| LaFe$_{10.7}$Co$_{0.8}$Si$_{1.5}$ | 285 | 7.0 | 13.5 | [34] |
| LaFe$_{10.98}$Co$_{0.22}$Si$_{1.8}$ | 242 | 6.3 | 11.5 | [11] |
| La$_{0.8}$Pr$_{0.2}$Fe$_{10.7}$Co$_{0.8}$Si$_{1.5}$ | 280 | 7.2 | 13.6 | [34] |
| La$_{0.6}$Pr$_{0.4}$Fe$_{10.7}$Co$_{0.8}$Si$_{1.5}$ | 274 | 7.4 | 14.2 | [34] |
| La$_{0.5}$Pr$_{0.5}$Fe$_{10.7}$Co$_{0.8}$Si$_{1.5}$ | 272 | 8.1 | 14.6 | [34] |
| La$_{0.5}$Pr$_{0.5}$Fe$_{10.5}$Co$_{1.0}$Si$_{1.5}$ | 295 | 6.0 | 11.7 | [33] |
| La$_{0.7}$Nd$_{0.3}$Fe$_{10.7}$Co$_{0.8}$Si$_{1.5}$ | 280 | 7.9 | 15.0 | [54] |
| LaFe$_{11.12}$Co$_{0.71}$Al$_{1.17}$ | 279 | 4.6 | 9.1 | [11] |
| La(Fe$_{0.98}$Co$_{0.02}$)$_{11.7}$Al$_{1.3}$ | 198 | 5.9 | 10.6 | [11] |
| Gd | 293 | 5.0 | 9.7 | [34] |

**Figure captions**

Figure 1. a) Curie temperature $T_C$, lattice constant $a$ and b) magnetic moment of Fe atoms as a function of Si content x for $LaFe_{13-x}Si_x$. Data indicated by the open circles were obtained from Ref. 19.

Figure 2. a) Magnetization isotherms and b) the Arrott plots of $LaFe_{11.4}Si_{1.6}$ on field increase and decrease. Temperature step is 2 K in the region of 200–230 K, and 5 K in 165–200 K and 230–265 K, and c) entropy change $\Delta S$ as a function of temperature for $LaFe_{11.6}Si_{1.6}$ and $LaFe_{10.4}Si_{2.6}$ (Ref. 12).

Figure 3. a) Lattice parameter $a$ and b) entropy change $\Delta S$ as a function of temperature under a field change of 0–5 T for $LaFe_{13-x}Si_x$. Inset plot shows $\Delta S$ as a function of Si content for $LaFe_{13-x}Si_x$ under a field change of 0–2 T (Ref. 24).

Figure 4. Observed (dot), calculated (line) neutron diffraction patterns and their difference for $LaFe_{11.4}Si_{1.6}$ at T = 2, 191 and 300 K (Ref. 25).

Figure 5. a) Mössbauer spectra of $LaFe_{11.7}Si_{1.3}$ at 190 K and b) Mössbauer spectra of $LaFe_{11.0}Si_{2.0}$ at 240 K in various external magnetic fields (Ref. 29). The dotted lines are paramagnetic subspectra.

Figure 6. a) Temperature dependence of entropy change $\Delta S$ for a) $La(Fe_{1-x}Co_x)_{11.9}Si_{1.1}$ (Ref. 32) and b) $La(Fe_{1-x}Mn_x)_{11.7}Si_{1.3}$ (Ref. 40) under field changes of 0–2 T and 0–5 T. Inset plot is the $\Delta S$ of $La(Fe_{1-x}Co_x)_{11.9}Si_{1.1}$ with x=0.06 compared with that of Gd and $Gd_5Si_2Ge_2$ for a field change of 0–5 T (Ref. 31).

Figure 7. a) Curie temperature $T_C$ and b) entropy change $\Delta S$ as a function of $R$ concentration for $La_{1-x}R_xFe_{11.5}Si_{1.5}$ ($R$ = Ce, Pr and Nd) (Ref. 44).

Figure 8. Temperature dependence of entropy change $\Delta S$ for a) $LaFe_{11.5}Si_{1.5}H_\delta$ under a field change of 0–5 T (Ref. 48), b) $La(Fe_{1-x}Mn_x)_{11.7}Si_{1.3}H_\delta$ under field changes of 0–2 T and 0–5 T (Ref. 49) and c) $LaFe_{11.6}Si_{1.4}C_\delta$ under a field change of 0–5 T (Ref. 50).

Figure 9. Curie temperature $T_C$ as a function of lattice constant for the hydrogenised and pressed $LaFe_{11.5}Si_{1.5}$ and Ce-doped $LaFe_{11.5}Si_{1.5}$. The open circles show the pure volume-effects while the solid circles the as-detected $T_C$ under pressures. Solid lines are guides for the eyes (Ref. 55).

Figure 10. Phase volume dependence of the Curie temperature $T_C$ for $La_{0.7}R_{0.3}Fe_{11.5}Si_{1.5}$ (solid symbols). The $T_C$-V relation of the $LaFe_{13-x}Si_x$ compounds is also presented for comparison (open circles). Inset plot shows the $\Delta T_C/T_C$-x relation, where $\Delta T_C$ is the difference of the Curie temperatures of the La$_{1-x}$R$_x$Fe$_{11.5}$Si$_{1.5}$ and LaFe$_{11.5}$Si$_{1.5}$ compounds under the same phase volume. Solid lines are guides for the eye (Ref. 56).

Figure 11. a) Magnetization isotherms measured by ascending magnetic field, b) temperature-dependent entropy change calculated from Maxwell relation (MR) and heat capacity (HC) for La$_{0.7}$Pr$_{0.3}$Fe$_{11.5}$Si$_{1.5}$, and c) a schematic diagram showing the calculation of entropy change when stepwise magnetic behaviors occur (Ref. 58).

Figure 12. Magnetization isotherms of La$_{1-x}$Pr$_x$Fe$_{11.5}$Si$_{1.5}$ (x = a) 0, b) 0.2 and c) 0.4) in the field ascending and descending processes (Ref. 59), and temperature-dependent hysteresis loss of d) La$_{1-x}$Nd$_x$Fe$_{11.5}$Si$_{1.5}$ (x = 0, 0.1, 0.2 and 0.3), e) La$_{1-x}$Pr$_x$Fe$_{11.5}$Si$_{1.5}$ (x = 0.1, 0.2, 0.3, 0.4 and 0.5) (Ref. 44) and f) La$_{0.5}$Pr$_{0.5}$Fe$_{11.5}$Si$_{1.5}$C$_\delta$ ($\delta$=0 and 0.3) (Ref. 53), respectively.

Figure 13. Relation between entropy change and hysteresis loss for LaFe$_{13-x}$Si$_x$-based compounds with different compositions.

Figure 14. a) Adiabatic temperature change $\Delta T_{ad}$ as a function of temperature obtained by direct measurements under a field change from 0 to 1.4 T, b) $\Delta T_{ad}$ as a function of applied magnetic field at different temperatures for LaFe$_{11.7}$Si$_{1.3}$ (Refs. 24 and 69), c) $\Delta T_{ad}$ calculated from the heat capacity measurements as a function of temperature and d) $\Delta T_{ad}$ as a function of Si content under different fields for LaFe$_{13-x}$Si$_x$.

Figure 15. Entropy change $\Delta S$ of La(Fe$_{1-x}$Co$_x$)$_{11.83}$Al$_{1.17}$ (x=0.06 and 0.08), compared with that of Gd for magnetic field changes of 0–2 and 0–5T (Ref. 76). Solid lines show the theoretical results calculated in the molecular field approximation for a field changing from 0 to 5 T.

Figure 16. Temperature dependence of entropy change $\Delta S$ for a) polycrystalline Ni$_{51.5}$Mn$_{22.7}$Ga$_{25.8}$ (Ref. 82) and b) single crystal Ni$_{52.6}$Mn$_{23.1}$Ga$_{24.3}$ (Ref. 84) under different field changes.

Figure 17. a) Magnetization isotherms and b) entropy change $\Delta S$ as a function of temperature for $R_6$Co$_2$Si$_3$ ($R$ = Pr, Gd and Tb) (Refs. 99, 100 and 101).

Figure 18. a) Entropy change $\Delta S$ and b) adiabatic temperature change $\Delta T_{ad}$ as a function of temperature for CdCr$_2$S$_4$ under different field changes (Ref. 109).

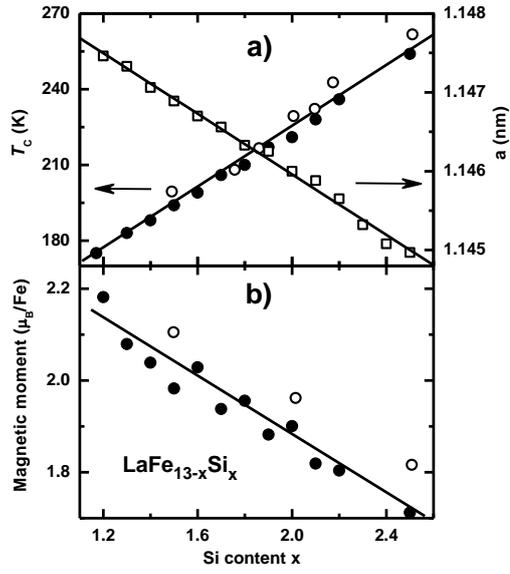

Figure 1 by Shen et al.

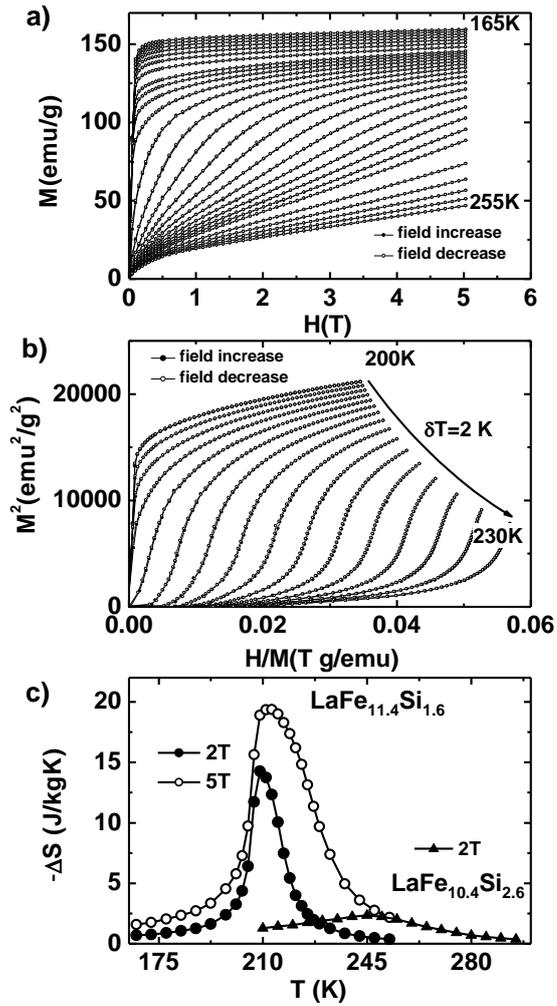

Figure 2 by Shen et al.

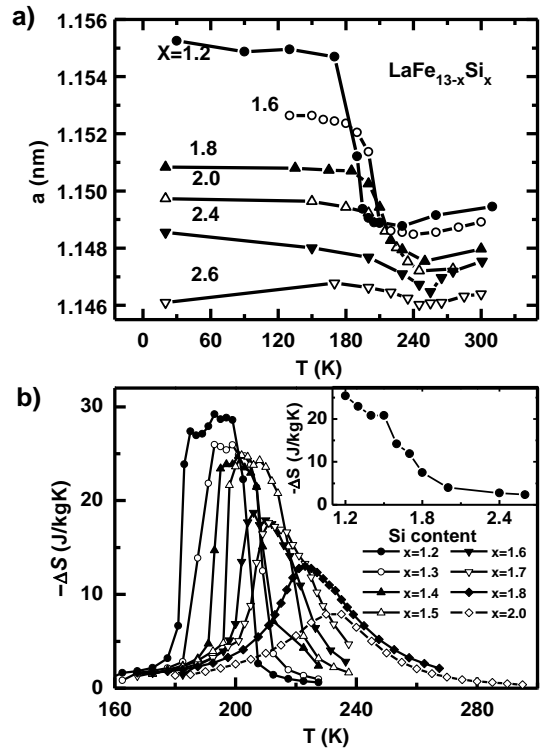

Figure 3 by Shen et al.

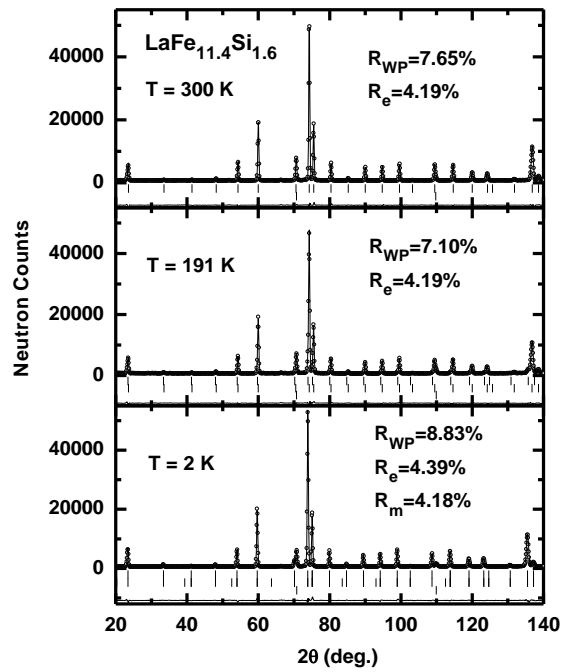

Figure 4 by Shen et al.

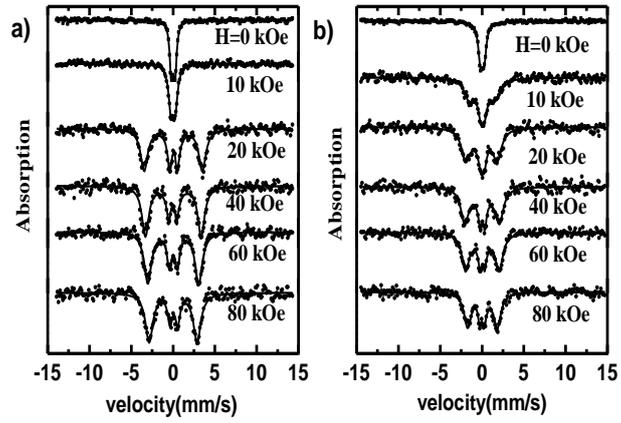

Figure 5 by Shen et al.

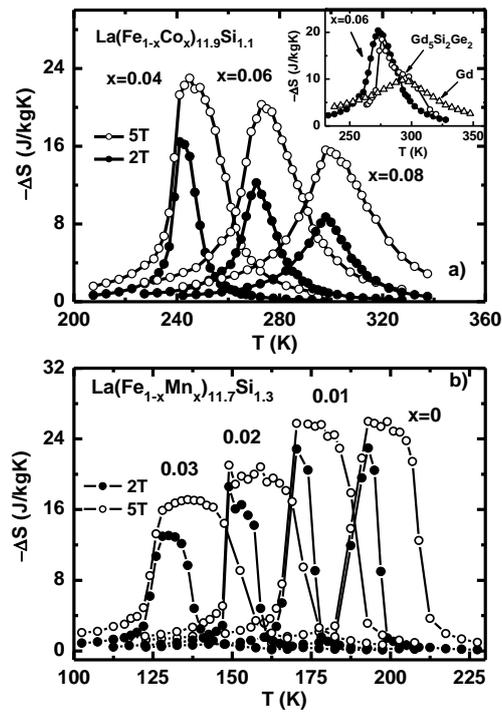

Figure 6 by Shen et al.

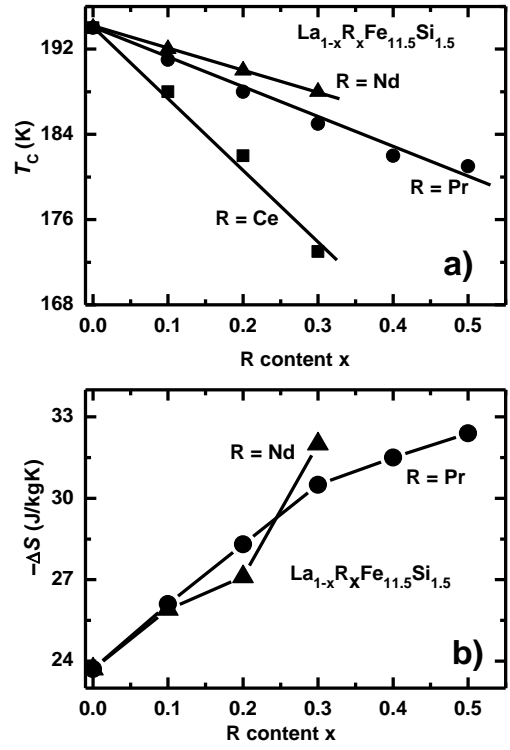

Figure 7 by Shen et al.

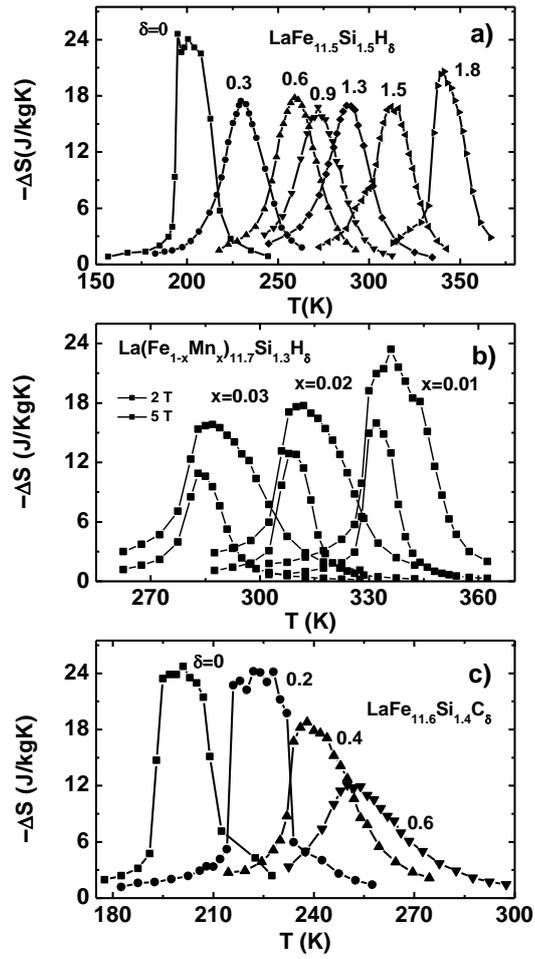

Figure 8 by Shen et al.

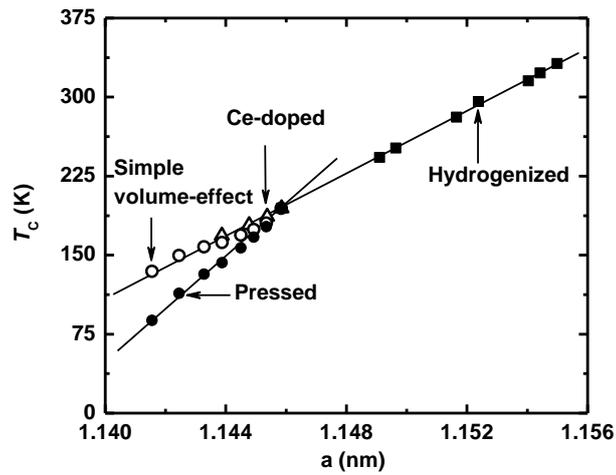

Figure 9 by Shen et al.

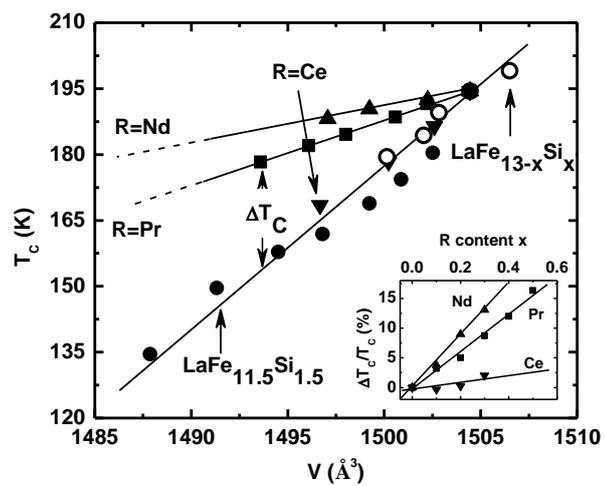

Figure 10 by Shen et al.

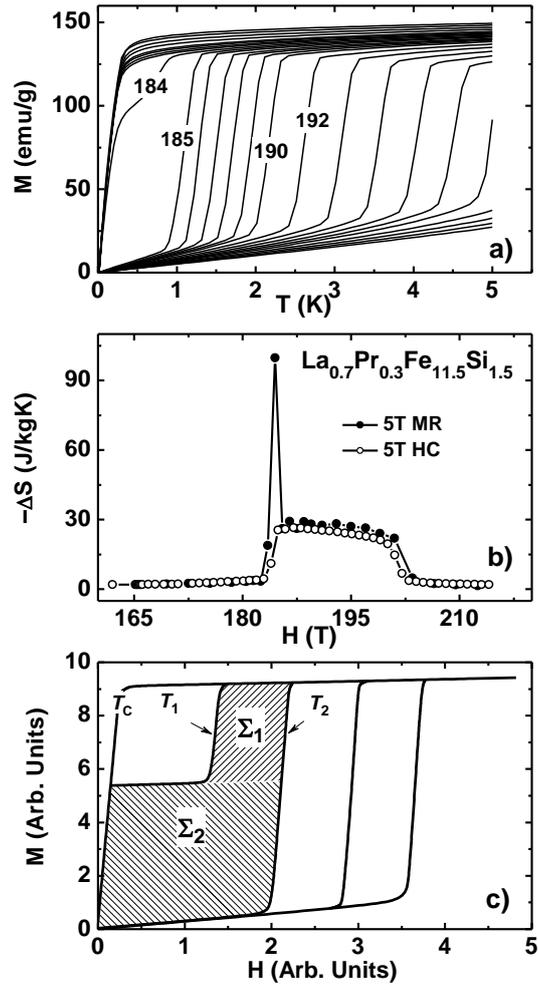

Figure 11 by Shen et al.

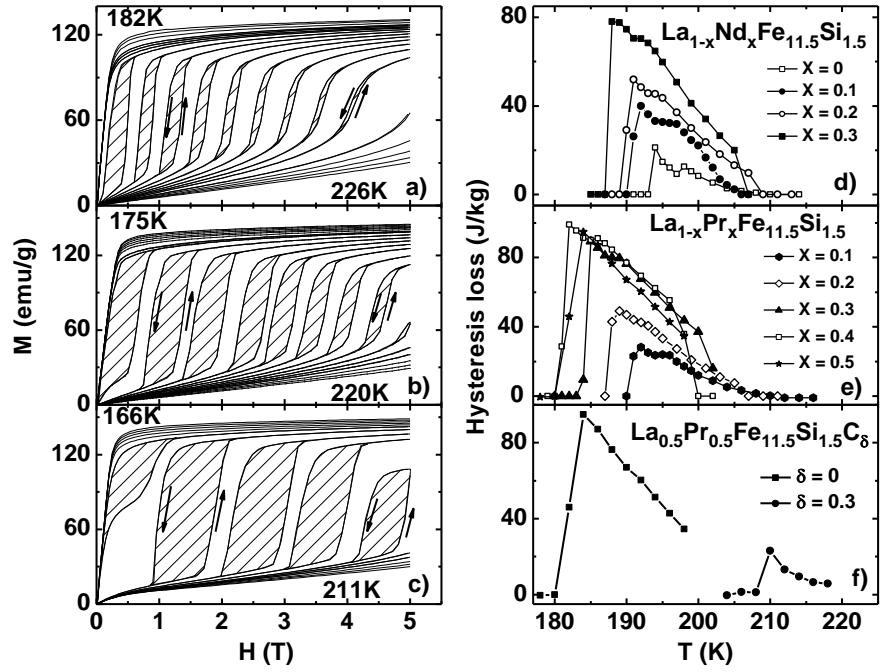

Figure 12 by Shen et al.

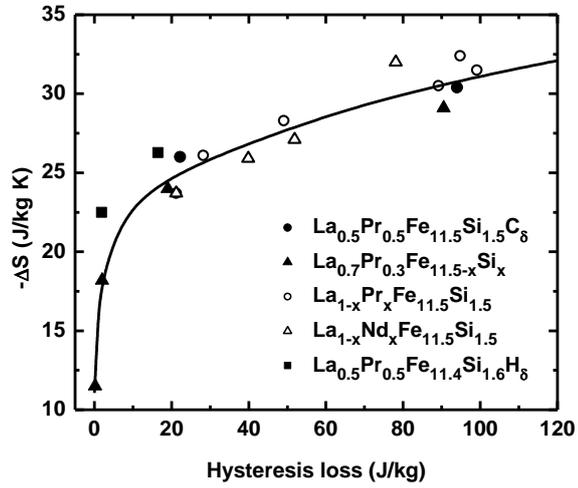

Figure 13 by Shen et al.

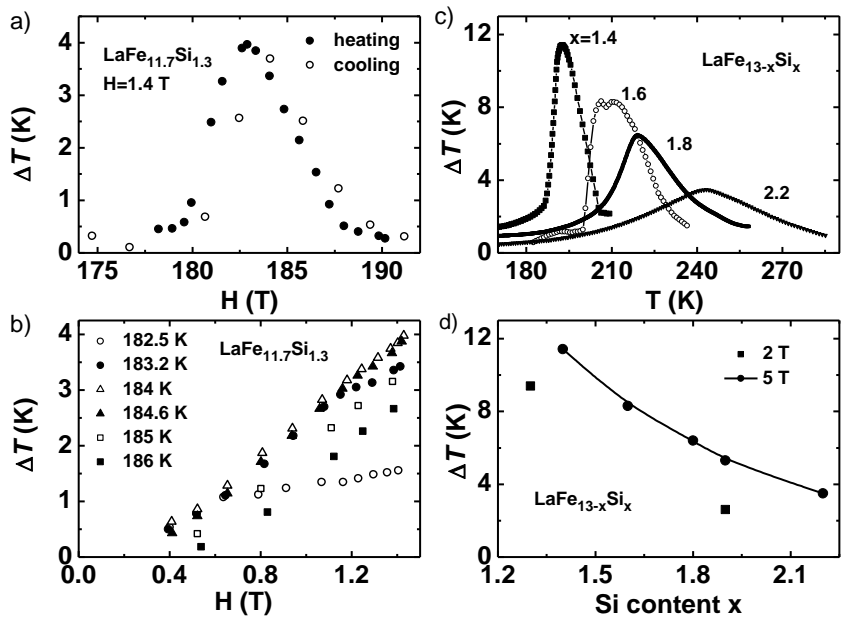

Figure 14 by Shen et al.

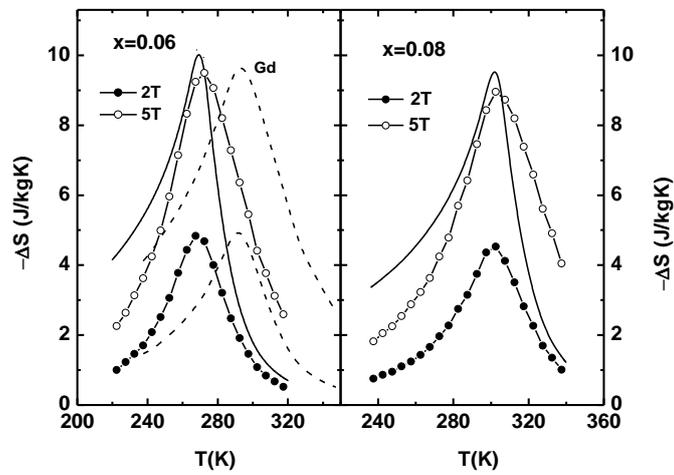

Figure 15 by Shen et al.

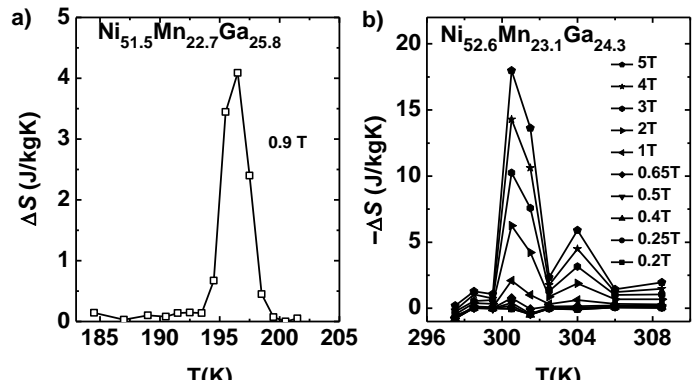

Figure 16 by Shen et al.

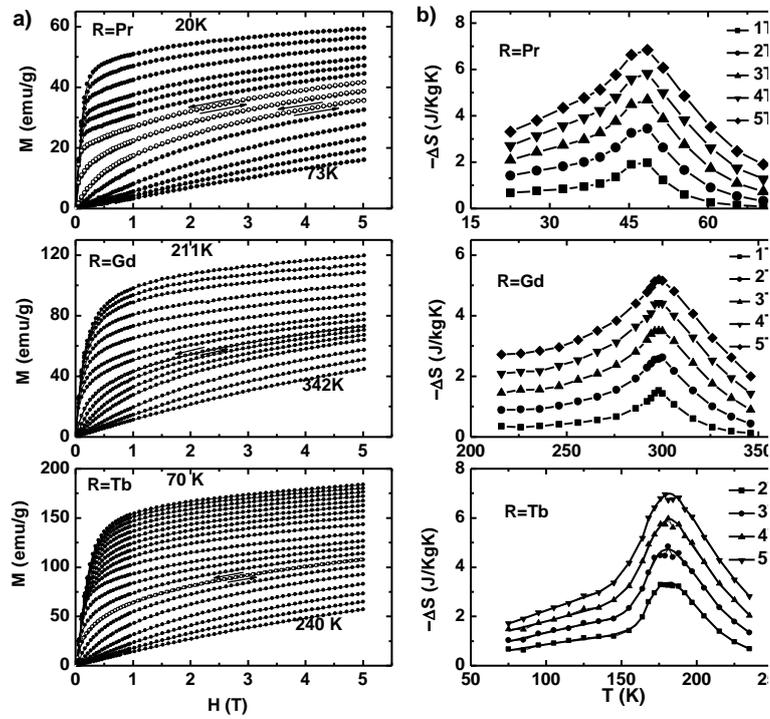

Figure 17 by Shen et al.

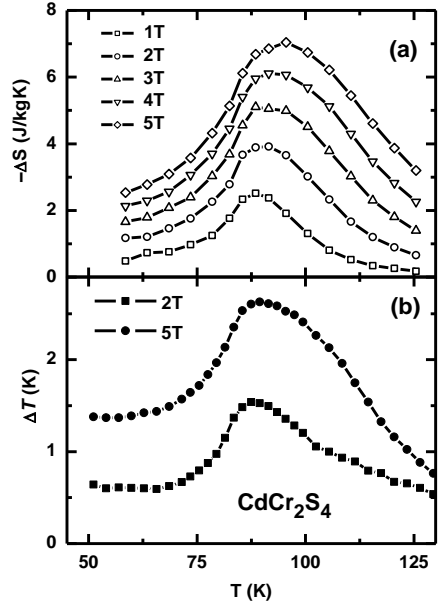

Figure 18 by Shen et al.

Table 1 Magnetic transition temperature $T_C$ and isothermal entropy change $|\Delta S|$ for $LaFe_{13-x}Si_x$ and related compounds

| compounds | $T_C$ (K) | $-\Delta S$ (J/kg K) 0–2 T | $-\Delta S$ (J/kg K) 0–5 T | Refs. |
|---|---|---|---|---|
| $LaFe_{11.83}Si_{1.17}$ | 175 | 21.2 | 27.8 | This work |
| $LaFe_{11.8}Si_{1.2}$ |  | 25.4 | 29.2 | This work |
| $LaFe_{11.7}Si_{1.3}$ | 183 | 22.9 | 26.0 | This work |
| $LaFe_{11.6}Si_{1.4}$ | 188 | 20.8 | 24.7 | This work |
| $LaFe_{11.5}Si_{1.5}$ | 194 | 20.8 | 24.8 | This work |
| $LaFe_{11.5}Si_{1.5}$ | 194 | 21.0 | 23.7 | [44] |
| $LaFe_{11.5}Si_{1.5}$ | 195 | 21.9 | 24.6 | [48] |
| $LaFe_{11.4}Si_{1.6}$ | 199 | 14.2 | 18.7 | This work |
| $LaFe_{11.4}Si_{1.6}$ | 209 | 14.3 | 19.3 | [11] |
| $LaFe_{11.4}Si_{1.6}$ | 208 | 14.3 | 19.4 | [12] |
| $LaFe_{11.3}Si_{1.7}$ | 206 | 11.9 | 17.6 | This work |
| $LaFe_{11.2}Si_{1.8}$ | 210 | 7.5 | 13.0 | This work |
| $LaFe_{11.0}Si_{2.0}$ | 221 | 4.0 | 7.9 | This work |
| $LaFe_{10.6}Si_{2.4}$ |  | 2.8 |  | [24] |
| $LaFe_{10.4}Si_{2.6}$ |  | 2.3 |  | [14] |
| $La_{0.9}Pr_{0.1}Fe_{11.5}Si_{1.5}$ | 191 | ~24 | 26.1 | [44] |
| La$_{0.8}$Pr$_{0.2}$Fe$_{11.5}$Si$_{1.5}$ | 188 | ~26 | 28.3 | [44] |
| La$_{0.7}$Pr$_{0.3}$Fe$_{11.5}$Si$_{1.5}$ | 185 | ~28 | 30.5 | [44] |
| La$_{0.6}$Pr$_{0.4}$Fe$_{11.5}$Si$_{1.5}$ | 182 | ~29 | 31.5 | [44] |
| La$_{0.5}$Pr$_{0.5}$Fe$_{11.5}$Si$_{1.5}$ | 181 | ~30 | 32.4 | [44] |
| La$_{0.9}$Nd$_{0.1}$Fe$_{11.5}$Si$_{1.5}$ | 192 | ~23 | 25.9 | [44] |
| La$_{0.8}$Nd$_{0.2}$Fe$_{11.5}$Si$_{1.5}$ | 190 | ~24 | 27.1 | [44] |
| La$_{0.7}$Nd$_{0.3}$Fe$_{11.5}$Si$_{1.5}$ | 188 | ~29 | 32.0 | [44] |
| La$_{0.7}$Ce$_{0.3}$Fe$_{11.5}$Si$_{1.5}$ | 173 | | 23.8 | [44] |
| La$_{0.7}$Pr$_{0.3}$Fe$_{11.4}$Si$_{1.6}$ | 190 | 25.4 | 28.2 | [38] |
| La$_{0.7}$Pr$_{0.3}$Fe$_{11.2}$Si$_{1.8}$ | 204 | 14.4 | 19.4 | [38] |
| La$_{0.7}$Pr$_{0.3}$Fe$_{11.0}$Si$_{2.0}$ | 218 | 6.2 | 11.4 | [38] |
| LaFe$_{11.5}$Si$_{1.5}$H$_{0.3}$ | 224 | | 17.4 | [48] |
| LaFe$_{11.5}$Si$_{1.5}$H$_{0.6}$ | 257 | | 17.8 | [48] |
| LaFe$_{11.5}$Si$_{1.5}$H$_{0.9}$ | 272 | | 16.9 | [48] |
| LaFe$_{11.5}$Si$_{1.5}$H$_{1.3}$ | 288 | 8.4 | 17.0 | [45,48] |
| LaFe$_{11.5}$Si$_{1.5}$H$_{1.5}$ | 312 | | 16.8 | [48] |
| LaFe$_{11.5}$Si$_{1.5}$H$_{1.8}$ | 341 | | 20.5 | [48] |
| La(Fe$_{0.99}$Mn$_{0.01}$)$_{11.7}$Si$_{1.3}$H$_\delta$ | 336 | 16.0 | 23.4 | [49] |
| La(Fe$_{0.98}$Mn$_{0.02}$)$_{11.7}$Si$_{1.3}$H$_\delta$ | 312 | 13.0 | 17.7 | [49] |
| La(Fe$_{0.97}$Mn$_{0.03}$)$_{11.7}$Si$_{1.3}$H$_\delta$ | 287 | 11.0 | 15.9 | [49] |
| LaFe$_{11.5}$Si$_{1.5}$C$_{0.2}$ | 225 | 18.0 | 22.8 | [51] |
| LaFe$_{11.5}$Si$_{1.5}$C$_{0.5}$ | 241 | 7.4 | 12.7 | [51] |
| La$_{0.5}$Pr$_{0.5}$Fe$_{11.5}$Si$_{1.5}$C$_{0.3}$ | 211 | 25.2 | 27.6 | [53] |
| La(Fe$_{0.96}$Co$_{0.04}$)$_{11.9}$Si$_{1.1}$ | 243 | 16.4 | 23.0 | [32] |
| La(Fe$_{0.94}$Co$_{0.06}$)$_{11.9}$Si$_{1.1}$ | 274 | 12.2 | 19.7 | [32] |
| La(Fe$_{0.92}$Co$_{0.08}$)$_{11.9}$Si$_{1.1}$ | 301 | 8.7 | 15.6 | [32] |
| LaFe$_{11.2}$Co$_{0.7}$Si$_{1.1}$ | 274 | | 20.3 | [31] |
| LaFe$_{10.7}$Co$_{0.8}$Si$_{1.5}$ | 285 | 7.0 | 13.5 | [34] |
| LaFe$_{10.98}$Co$_{0.22}$Si$_{1.8}$ | 242 | 6.3 | 11.5 | [11] |
| La$_{0.8}$Pr$_{0.2}$Fe$_{10.7}$Co$_{0.8}$Si$_{1.5}$ | 280 | 7.2 | 13.6 | [34] |
| La$_{0.6}$Pr$_{0.4}$Fe$_{10.7}$Co$_{0.8}$Si$_{1.5}$ | 274 | 7.4 | 14.2 | [34] |
| La$_{0.5}$Pr$_{0.5}$Fe$_{10.7}$Co$_{0.8}$Si$_{1.5}$ | 272 | 8.1 | 14.6 | [34] |
| La$_{0.5}$Pr$_{0.5}$Fe$_{10.5}$Co$_{1.0}$Si$_{1.5}$ | 295 | 6.0 | 11.7 | [33] |
| La$_{0.7}$Nd$_{0.3}$Fe$_{10.7}$Co$_{0.8}$Si$_{1.5}$ | 280 | 7.9 | 15.0 | [54] |
| LaFe$_{11.12}$Co$_{0.71}$Al$_{1.17}$ | 279 | 4.6 | 9.1 | [11] |
| La(Fe$_{0.98}$Co$_{0.02}$)$_{11.7}$Al$_{1.3}$ | 198 | 5.9 | 10.6 | [11] |
| Gd | 293 | 5.0 | 9.7 | [34] |